\documentclass[journal]{IEEEtran}
\IEEEoverridecommandlockouts
\usepackage{cite}
\usepackage{amsmath,amssymb,amsfonts}
\usepackage{algorithmic}
\usepackage{graphicx}
\usepackage{textcomp}
\usepackage{xcolor}
\def\BibTeX{{\rm B\kern-.05em{\sc i\kern-.025em b}\kern-.08em
    T\kern-.1667em\lower.7ex\hbox{E}\kern-.125emX}}
\usepackage{amsfonts}
\usepackage{amsmath}
\usepackage{amsthm}

\usepackage{float}
\usepackage{graphicx}
\usepackage{epstopdf}
\usepackage{multirow}
\usepackage{caption}
\usepackage{subcaption}
\usepackage{ragged2e}
\usepackage{comment}
\usepackage{enumitem}
\usepackage[colorlinks =True,
            linkcolor  =blue,
            anchorcolor=green,
            citecolor  =blue]{hyperref} 
\usepackage[ruled,vlined]{algorithm2e}

\theoremstyle{plain}

\newtheorem*{theorem*}{Theorem}

\def\D{{\mathcal D}}

\def\R{{\mathbb  R}}
 
\def\U{{\mathbf  U}}
\def\T{{\mathbf  Z}}
\def\W{{\mathbf  W}}
\def\V{{\mathbf  V}}
\def\N{{\mathbb  N}}
\def\bu{\mathbf{u}}
\def\bv{\mathbf{v}} 
\def\bw{\mathbf{w}}
\def\bz{\mathbf{z}}
\def\bx{\mathbf{x}} 
\def\by{\mathbf{y}}

\def\bta{{\mathbf z}}

\def\bpi{{\boldsymbol  \pi}}

\def\P{{\boldsymbol  \Pi}}
\def\eqspace{\arraycolsep=1.5pt\def\arraystretch}
\graphicspath{ {./images/} }

\begin{document}
\title{
 New Environment Adaptation with Few Shots for OFDM Receiver and mmWave Beamforming }

\author{Ouya Wang, Shenglong Zhou, and Geoffrey Ye Li
\thanks{Ouya Wang and Geoffrey Ye Li are with the Department of Electrical and Electronic Engineering, Faculty of Engineering, Imperial College London, London, U.K. (e-mail: \{ouya.wang20, geoffrey.li\}@imperial.ac.uk).}%
\thanks{Shenglong Zhou (corresponding author) is with the School of Mathematics and Statistics, Beijing Jiaotong University, Beijing, China. (e-mail: slzhou2021@163.com)}}


\maketitle
\begin{abstract}  
Few-shot learning (FSL) enables adaptation to new tasks with only limited training data. In wireless communications, channel environments can vary drastically; therefore, FSL techniques can quickly adjust transceiver accordingly. In this paper, we develop two FSL frameworks that fit in wireless transceiver design. Both frameworks are base on optimization programs that can be solved by well-known algorithms like the inexact alternating direction method of multipliers (iADMM) and the inexact alternating direction method (iADM). As examples, we demonstrate how the proposed two FSL frameworks are used for the OFDM receiver and beamforming (BF) for the millimeter wave (mmWave) system. The numerical experiments confirm their desirable performance in both applications compared to other popular approaches, such as transfer learning (TL) and model-agnostic meta-learning.   

 
\end{abstract}
\begin{IEEEkeywords}
Few-shot learning, iADMM, iADM, wireless communication, hyper network
\end{IEEEkeywords}

\section{Introduction}\label{sec::intro}
Deep learning (DL), as a transformative technology, is capable of enhancing the performance of wireless communication systems significantly. Its ability to automatically learn intricate features from data has opened new lines for various applications, such as improved signal processing \cite{ye2017power}, interference mitigation \cite{xia2020mimo}, and resource allocation \cite{liang2017resource}. In particular, DL-based methods have found extensive use in the realm of physical layer communications \cite{qin2019deep}. For example, in \cite{ye2017power}, a deep neural network (DNN) is adopted to jointly optimize channel estimation and signal detection in a data-driven manner. A model-driven DL is proposed in \cite{he2019model}, where a trainable framework is coupled with the orthogonal approximate message passing detector for massive multiple-input multiple-output detection. These methods perform relatively well under known/trained environments but may suffer from performance degradation when deploying trained models in unseen/new environments.

DL techniques can be also used in beam prediction. In the mmWave system, DL can exploit additional contextual information, such as receiver locations and surrounding obstacles in mobile networks, to improve predictions accuracy \cite{qin2019deep}. By leveraging DL to acquire insights on received signals at base stations (BS), the developed solution in \cite{alkhateeb2018deep} makes full use of wide-coverage and low-latency coordinated BF gains with minimal coordination overhead. The BF neural network in \cite{lin2019beamforming} is proposed to optimizes the beamformer to maximize spectral efficiency with imperfect channel state information. Once again, these DL-based approaches outperform conventional BF algorithms but their performance degrades in quickly changing environments.

DL-based methods face at least three challenges when used in wireless communication systems. First, the system frequently undergoes rapid and unpredictable environmental changes. Acquiring and annotating extensive data to account for these rapid changes becomes impractical and resource-intensive, hindering real-time adaptation to new environments. Secondly, when encountering a new environment, the amount of available data is typically limited, placing significant strain on conventional DL approaches that usually demand vast labeled datasets for effective training. Finally, data distributions across various wireless scenarios exhibit distinct characteristics, which impedes the universal applicability of a single model across diverse domains.

Hence, effectively obtaining valid information from a small number of samples becomes essential to satisfy the demand for real-world wireless communications. FSL is an emerging paradigm that addresses these challenges by producing creative work on data, models, and algorithms. According to \cite{song2023comprehensive}, FSL techniques can be divided into data augmentation, multimodal learning, meta-learning \cite{thrun1998learning,simeone2020learning,wang2022learn}, and transfer learning (TL) \cite{liu2019overcoming,pati2020deep,van2022transfer}. One critical issue of FSL in wireless communications is that the data from different environments are heterogeneous and the samples of a new environment are small. Despite the similarities shared by different environments, each still has its unique characteristics. How do we extract these common and unique characteristics? Moreover, when a new environment comes with limited samples, how can we leverage prior experience from known environments to enable the wireless system to adapt to the new environment? Another challenge lies in the slow convergence during training. Given that the system needs to learn from a wide range of different environments to accumulate experience, conventional deep learning optimizers struggle to efficiently discover the optimal shared and specific parameters within limited epochs. This inevitably leads to extensive training times and high computational resource requirements. How do we develop an optimal algorithm based on conventional DL tools to accelerate the training convergence speed in this scenario? 

\subsection{FSL Adaptation Problem}
Suppose that we are given ${n}$ datasets  (representing ${n}$ previous environments), denoted as $\{\D_0,\D_1,\cdots,\D_{n-1}\}$, with $\D_i:=\{(\bx_i^{t},\by_i^{t}): t=1,2,\cdots,d_i \}, i=0,1,\cdots,n-1$, where ${\bx_i^t}$ and ${\by_i^t}$ stand for the features/inputs and labels/outputs and $d_i $ is the cardinality of $\D_i $. Now we encounter a new environment (i.e., the $n$th environment) with a small dataset $\D_n:=\{(\bx_n^{t},\by_n^{t}): i=1,2,\cdots,d_n\}$, where $d_n\ll \min\{d_0,$ $d_1,\cdots,d_{n-1}\}$. All these datasets share some common patterns, in addition to their own unique characteristics.
So the question is how can we leverage the prior knowledge from the $n$ previous environments and the limited samples from the new (the $n$th) environment to make a decision, particularly in determining ${\by_n}$ when new data ${\bx_n}$ emerges?

\subsection{Contribution}
The main contributions of the paper are threefold. To begin with, we propose two new learning schemes, to fulfill FSL tasks in wireless communications. Both schemes provide versatile and adaptable methods that can be applied to a broad range of learning tasks, making them promising for large-scale applications. Moreover, we employ the inexact alternating direction method of multipliers (iADMM) and the inexact alternating direction method (iADM) to tackle the non-convex optimization problems involving neural networks. Our numerical results have shown that they exhibit faster convergence compared to the conventional DL optimizers, such as stochastic gradient descent (SGD) and root mean-squared propagation (RMSProp). Furthermore, we have designed a novel transformer-based OFDM receiver architecture and a DL-based mmWave BF prediction system as wireless application examples. They are augmented with a hyper-network that employs few-shot samples to generate parameters, enabling the system to adapt to various environmental conditions.

We would like to point out that a portion of our work has been previously published in \cite{wang2023efficient}. In this paper, we have made two additional contributions: Firstly, we have introduced the framework of online adaptation FSL, enabling rapid adaptation to new environments. To testify its efficiency, we develop a DL-based mmWave BF prediction system. Secondly, we have incorporated second-order gradients into the iADMM algorithm, resulting in much faster convergence during offline training compared to RMSProp. 

\subsection{Organization and Notation}
The rest of the paper is organized as follows. In Section \ref{EA_OFDM}, we propose the effective adaptation (EA) framework and investigate its application in OFDM receiver design. In Section \ref{OA:BF}, we introduce the online adaptation (OA) framework and its application in mmWave BF prediction. Finally, conclusion remarks are provided in Section \ref{sec:conclusion}.
 
We end this section by introducing some notations used in this paper. We denote $\N:=\{0,1,\cdots,n-1\}$. Here $:=$ means define. Let ${\|\cdot\|}$ be the Euclidean norm for vectors and the spectral norm for matrices. In the sequel, superscript $t$ represents the index of a sample (e.g. $\bx_i^t$), superscript $\ell$ stands for the iteration number (e.g., $\bw_i^{\ell}$ or $\bz_i^{\ell}$), and subscript $i$ is an entry of $\N$. Moreover, we denote several matrices as follows:
\begin{eqnarray*}\eqspace{1.5}
\begin{array}{lll}
\W&:=&(\bw_0,\bw_1\cdots,\bw_{n-1}),\\
\W^{\ell}&:=&(\bw_0^{\ell},\bw_1^{\ell}\cdots,\bw_{n-1}^{\ell}),\\
\W^*&:=&(\bw_0^*,\bw_1^*\cdots,\bw_{n-1}^*).
\end{array}\end{eqnarray*}
Similar rules are also applied to other matrices, such as $\V,\P,\U$, and $\T$.


\section{Effective Adaptation with Application in OFDM Receiver}\label{EA_OFDM}
In this section, we first introduce the EA learning schemes from a mathmatical perspective, and then develop a corresponding algorithm for EA learning. Finally, we present experiment for the EA application to the OFDM receiver, which includes system settings, experiment configurations, adaptation performance and comparison with some state-of-the-art (SOTA) alternatives.   

\subsection{Problem Formulation}\label{EA_Problem}
\begin{figure}[h]
\centering
\includegraphics[scale=.485]{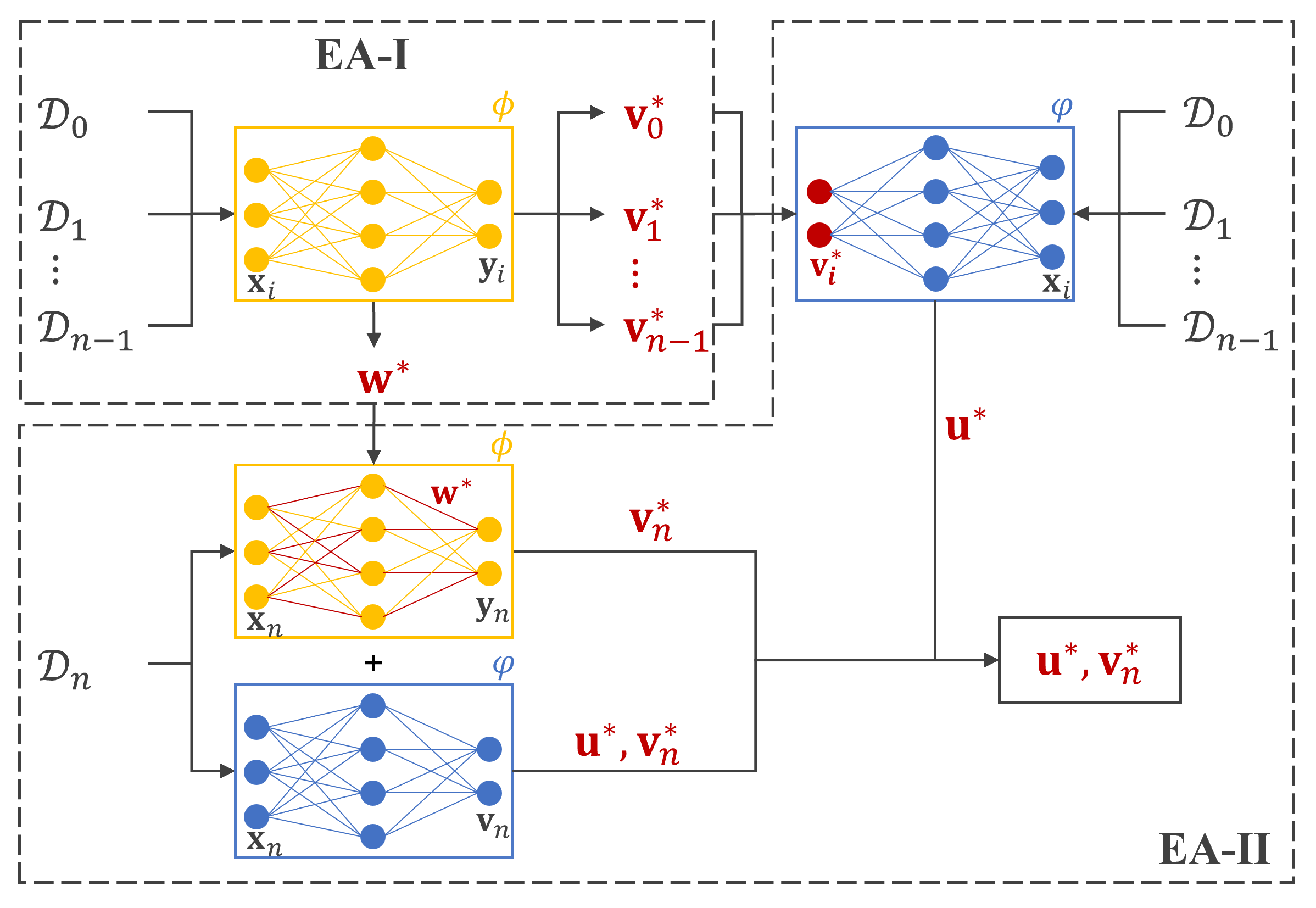}
\caption{Structure of EA.}\label{fig:EA}
\end{figure}

This approach comprises two phases, as shown in Fig. \ref{fig:EA}. The first phase learns knowledge from $n$ environments which together with the new environmental data helps the second phase gain knowledge of the new environment so as to make a decision.

{\bf EA-I: Learning from previous environments}  Let ${\bw}^*$ represent the shared parameter (corresponding to similar patterns) and ${\V^*:=( {\bv}_0^*,{\bv}_1^*,\cdots, {\bv}_{n-1}^*)}$ stand for ${n}$ particular parameters  (corresponding to ${n}$  unique characteristics) for the ${n}$ environments. To learn these parameters, we solve the following optimization problem,
\begin{eqnarray}\label{phase-i-opt}
\begin{array}{l}({\bw}^*,\V^*)
:=\underset{\bw,\V}{\rm argmin} ~\frac{1}{n} \sum_{i=0}^{n-1} f^\phi_i (\bw,\bv_i ),\end{array} 
\end{eqnarray}
where 
\begin{eqnarray}\label{obj-phi-n}
\begin{array}{l}f^\phi_i (\bw,\bv_i ):= \sum_{t=1}^{d_i } \ell\left(\phi(\bw,\bv_i ;\bx^{t}_i ),\by^{t}_i  \right),~~ i\in\N.\end{array}
\end{eqnarray}
Here, $\ell(\cdot,\cdot)\geq 0$ is a loss function and ${\phi(\cdot,\cdot;\bx)}$ can be regarded as a communication system realized by a neural network with input $\bx$. Solving the above problem gives rise to a relationship between the inputs and outputs as
\begin{eqnarray}\label{relationship-features-actions}
 \by_i  \approx \phi({\bw}^*,{\bv}_i ^*;\bx_i ), ~i\in\N.
\end{eqnarray}

 {\bf EA-II: Adaptation into the new environment} In phase EA-I, similar to the other environments, we can learn ${\bv}_n^*$ directly, which however may obtain undesirable results since the samples of the new environment are insufficient. Alternatively, we first try to find a shared parameter $\bu$ such that 
\begin{eqnarray}\label{relationship-hypernetwork}
\eqspace{1.5}
\begin{array}{llll}
{\bv}_i ^*&\approx& \varphi(\bu;\bx_i ^{t}),~& t=1,2,\cdots, d_i ,~~i\in\N,\\
{\bv}_n^*&\approx&\varphi(\bu;\bx_n^{t}),~& t=1,2,\cdots, d_n,
\end{array}
\end{eqnarray}

where  $\varphi(\cdot;\bx)$  is a hyper-network with input $\bx$. This aims at further digging out the relationship between input $\bx$ and output $\bv$. Then we leverage the new data to learn ${\bv}_n^*$. Overall, the second phase is to solve the following problem, 
\begin{eqnarray}\label{phase-ii-opt}
\eqspace{1.5}
\begin{array}{lll}
({\bu}^*,\bv_n^*)&:=&\underset{\bu,~\bv_n}{\rm argmin}~  \frac{1}{n}\sum_{i=0}^{n-1} f^\varphi_i (\bu,\bv_i ^*) \\
&+& f^\varphi_n(\bu,\bv_n) +\lambda f^\phi_n(\bw^*,\bv_n) ,
\end{array}
\end{eqnarray}
where 
\begin{eqnarray}\label{obj-phi-new}
\eqspace{1.75}
\begin{array}{lll}
f^\varphi_i (\bu,\bv_i )&:=& \sum_{t=1}^{d_i } \left \|\varphi(\bu;\bx^{t}_i )- {\bv}_i \right\|^2, ~i\in\N ,\\
f^\varphi_n(\bu,\bv_n)&:=& \sum_{t=1}^{d_n }\left \|\varphi(\bu;\bx^{t}_n)- {\bv}_n\right\|^2,\\
f^\phi_n(\bw,\bv_n)&:= &\sum_{t=1}^{d_n} \ell\left(\phi(\bw,\bv_n;\bx^{t}_n),\by^{t}_n \right).
\end{array}
\end{eqnarray}
Here, ${\lambda> 0}$ is a penalty parameter. One can observe that the objective function in \eqref{phase-ii-opt} consists of three parts. The first two parts are from \eqref{relationship-hypernetwork} to explore the relationship between $\bx$ and $\bv$, and the purpose of using the last term is similar to \eqref{phase-i-opt} that tries to reduce the training loss on the new data.

As shown in Fig. \ref{fig:EA}, two neural networks $\phi$ and $\varphi$ are used for the training. The former aims at learning shared parameter $\bw^*$ and unique parameters $\{\bv_i^*:i\in\N\}$ for $N$ previous environments and $\bv_n^*$ for the new environment. The latter aims at training $\bu^*$ and $\bv_n^*$.  

Finally, we use TL techniques to transfer the shared weights, together with learned parameter ${({\bu}^*,\bv_n^*)}$ into \eqref{relationship-features-actions}. Suppose we encounter data ${\bx_n^{\rm new}}$ in this new environment, then we can take a response by 
 \begin{eqnarray}\label{actions-new-environment}
 \by_n^{\rm new} = \phi\left(\bw^*,\bv^*_n;\bx_n^{\rm new}\right).
\end{eqnarray}

%

As shown in Fig. \ref{fig:EA}, data from all previous environments are reused in the process (i.e., EA-II) of adapting to the new environment, that is,  we reuse ${\{\D_i: i\in\N\}}$ to solve problem \eqref{phase-ii-opt}, which evidently needs more computational endeavours. This approach might be impractical for scenarios where different new environments are encountered, as we would solve problem \eqref{phase-ii-opt} for each new environment. Therefore, to overcome this drawback, in the following subsection, we develop another approach purely based on the new data when adapting prior experience to the new environment. 

\subsection{Solving problem (\ref{phase-i-opt})}
One can discern that problems \eqref{phase-i-opt} and \eqref{phase-ii-opt} involve two neural networks, i.e., $\phi$ and $\varphi$, and thus are non-convex in general. We note that the inexact ADMM (iADMM) has shown its popularity and ability in the recent several decades to solve some complex non-convex optimization. Therefore, we adopt it to address the optimization problems in Section \ref{EA_Problem}. We first equivalently rewrite problem \eqref{phase-i-opt}  as 
\begin{eqnarray}\label{phase-i-opt-eq}
\eqspace{1.5}
\begin{array}{r}
({\bw}^*,{\W}^*,\V^*):= {\rm argmin}_{(\bw,\W,\V)}  \frac{1}{n} \sum_{i=0}^{n-1} f^\phi_i (\bw_i ,\bv_i ) \\
{\rm s.t.}~ \bw_i =\bw, ~i\in\N .
\end{array}
\end{eqnarray}
The augmented Lagrange function of the above problem is
\begin{align*}
\eqspace{1.5}
\begin{array}{rrr}
L( \bw , \W, \V, \P) :=  \sum_{i=0}^{n-1} L_i (\bw, \bw_i , \bv_i , \bpi_i ),
\end{array}
\end{align*}
where $\P=(\bpi_0,\bpi_1,\cdots,\bpi_{n-1})$ are the Lagrange multipliers, 
\begin{align*}
\eqspace{1.5}
\begin{array}{l}
L_i (\bw , \bw_i , \bv_i , \bpi_i )\\
:=\frac{1}{n} f^\phi_i (\bw_i ,\bv_i ) + \langle \bpi_i , \bw_i -\bw \rangle
 +\frac{\sigma}{2}\|\bw_i -\bw\|^2,
\end{array}
\end{align*}
and $\sigma>0$ is a given constant. Then the framework of iADMM can be described as follows. Given $(\W^0,\V^0, \P^0)$, perform the following steps iteratively for $\ell=0,1,2,\cdots,$
\begin{align}\label{phase-i-ADMM}
\eqspace{1.5}
\begin{array}{r}
\bw^{\ell+1} =  {\rm argmin}~ L(\bw, \W^{\ell}, \V^{\ell}, \P^{\ell}) 
= \frac{1}{n}\sum_{i=0}^{n-1}(\bw_i ^{\ell}+\frac{\bpi_i ^{\ell}}{\sigma}) \\
(\bw_i ^{\ell+1} , \bv_i ^{\ell+1} ) \approx~ {\rm argmin}~ L_i (\bw^{\ell+1}, \bw_i , \bv_i , \bpi_i ^{\ell}),~i\in\N ,\\
 \bpi_i ^{\ell+1} = \bpi_i ^{\ell} + \sigma (\bw_i ^{\ell+1}-\bw^{\ell+1}),~i\in\N .
\end{array}
\end{align}
 One can observe that the critical step is to solve the second problem in \eqref{phase-i-ADMM}. To accelerate the computation, we aim to solve it inexactly by
  \begin{align}\label{phase-i-ADMM-inexact}
\eqspace{1.5}
\begin{array}{r}
(\bw_i ^{\ell+1} , \bv_i ^{\ell+1} )=  {\rm argmin}~ \langle ({\boldsymbol  \zeta}_i ^{\ell}, ~{\boldsymbol  \xi}_i ^{\ell}), (\bw_i , \bv_i )-(\bw^{\ell+1},\bv_i ^{\ell}) \rangle\\ 
 + \frac{\rho}{2}\|(\bw_i , \bv_i )-(\bw^{\ell+1},\bv_i ^{\ell})\|^2 \\ 
 +\langle \bpi_i ^{\ell}, \bw_i -\bw^{\ell+1} \rangle + 
\frac{\sigma}{2}\|\bw_i -\bw^{\ell+1}\|^2\\ 
= (
\bw^{\ell+1}-\frac{1}{\rho  + \sigma  }  ( {\boldsymbol  \zeta}_i ^{\ell}+\bpi_i ^{\ell}  ), ~\bv_i ^{\ell}   -\frac{1}{\rho }   {\boldsymbol  \xi}_i ^{\ell}),
 \end{array}
\end{align} 
where ${\rho>0}$  and ${({\boldsymbol  \zeta}_i ^{\ell},{\boldsymbol  \xi}_i ^{\ell})\in \frac{1}{n} \partial f^\phi_i (\bw^{\ell+1},\bv_i ^{\ell})}$. Here, notation    ${\partial f^\phi_i (\bw,\bv)}$, is the sub-differential of $f^\phi_i $ at ${(\bw,\bv)}$ which reduces to the gradient of $f^\phi_i $ if it is continuously differentiable.
One can replace  $\bw^{\ell+1}$ by $\bw_i ^{\ell}$ in the above problem to update ${(\bw_i ^{\ell+1} , \bv_i ^{\ell+1} )}$. However, this would degrade the convergence performance of the proposed algorithm as using  $\bw^{\ell+1}$  enables all ${\{\bw_i ^{\ell}:i\in\N \}}$ to converge to same value $\bw^{\ell+1}$, which makes it faster to satisfy the constraint in \eqref{phase-i-opt-eq}. Overall, we present these updates in Algorithm \ref{algorithm-ADMM}.

 \begin{algorithm}[!th]
\SetAlgoLined
{\noindent \justifying Initialize $(\W^0, \V^0, \P^0)$, $\rho>0$, and $\sigma>0$. Set $\ell = 0$.   }

\For{$\ell=0,1,2,\cdots $}{Update  $\bw^{\ell+1}$ by the first equation in \eqref{phase-i-ADMM}.
 
\For{$i \in\N$}{ Update  $(\bw_i ^{\ell+1} , \bv_i ^{\ell+1} )$ by \eqref{phase-i-ADMM-inexact}.\\
 Update  $\bpi_i ^{\ell+1}$ by the last equation in  \eqref{phase-i-ADMM}. 
 
 } 
} 
Return $\bw^{\ell}$ and $\{\bv_i ^{\ell}:i\in\N\}$.
\caption{iADMM solving problem (\ref{phase-i-opt}) \label{algorithm-ADMM}}
\end{algorithm}

\subsection{Solving problem (\ref{phase-ii-opt})}
For any $i\in\N $, we define the following function, 
\begin{eqnarray*}\label{notation-function-phase-II}
\eqspace{1.5}
\begin{array}{lll}
h_i (\bu_i , \bv_n)
:=  \frac{1}{n} \left(f^\varphi_i (\bu,\bv_i ^*) + f^\varphi_n(\bu,\bv_n) +\lambda f^\phi_n(\bw^*,\bv_n)\right)
\end{array}\end{eqnarray*}
By introducing $\bu_i =\bu, i\in\N $,  problem \eqref{phase-ii-opt} is equivalent to
 \begin{eqnarray*}\label{phase-ii-opt-eq}
 \eqspace{1.5}
\begin{array}{rl}
(\bu^*,\bv_n^*,\U^*) :=\underset{(\bu,\bv_n,\U)}{\rm argmin}~&  \sum_{i=0}^{n-1} h_i (\bu_i , \bv_n), \\ 
{\rm s.t.}~& \bu_i =\bu,~~i\in\N .
\end{array}
\end{eqnarray*}
 Similarly, for each $i\in\N$, we denote 
\begin{eqnarray*}\label{phase-ii-opt-Lag}
\eqspace{1.5}
\begin{array}{l}
H_i (\bu, \bv_n, \bu_i , \bz_i ) 
:=h_i (\bu_i ,\bv_n) + \langle \bta_i , \bu_i -\bu \rangle  
+ \frac{\mu}{2}\|\bu_i -\bu\|^2,
\end{array}
\end{eqnarray*}
where ${\mu>0}$ and ${\{\bta_i :i\in\N\}}$ are the Lagrange multipliers. Then given $(\U^0, \T^0)$, iADMM performs the following steps iteratively for $\ell=0,1,2,\cdots,$  
\begin{align}\label{phase-ii-ADMM} 
\eqspace{1.75}
\begin{array}{lll}
\bu^{\ell+1} &=&\frac{1}{n}\sum_{i=0}^{n-1}(\bu_i ^{\ell}+\frac{\bta_i ^{\ell}}{\mu}), \\ 
\bv_n^{\ell+1} &\approx& {\rm argmin}~\sum_{i=0}^{n-1} H_i (\bu^{\ell+1}, \bv_n , \bu_i ^{\ell}, \bta_i ^{\ell}),\\ 
\bu_i ^{\ell+1} &\approx& {\rm argmin}~ H_i (\bu^{\ell+1}, \bv_n^{\ell+1}, \bu_i , \bta_i ^{\ell}),~ i\in\N ,\\ 
 \bta_i ^{\ell+1} &=& \bta_i ^{\ell} + \mu (\bu_i ^{\ell+1}-\bu^{\ell+1}),~i\in\N .
\end{array}
\end{align}
Now let ${\nabla_{\bu} h_i (\bu,\bv )}$ and ${\nabla_{\bv} h_i (\bu,\bv )}$ be the partial gradients of $h_i $ with respect to $\bu$ and $\bv$, and $\eta>0$ and $\gamma>0$ be two given constants. Denote ${\boldsymbol\xi}_n^{\ell}:=\sum_{i\in\N} \nabla_{\bv}    h_i (\bu^{\ell+1},\bv_n^{\ell})~\text{and}~{\boldsymbol \zeta}_i ^{\ell}:= \nabla_{\bu}    h_i (\bu^{\ell+1},\bv_n^{\ell+1}).$ In order to accelerate the computation, we approximately update $\bv_n^{\ell+1} $  by 
 \begin{eqnarray} 
 \eqspace{1.5}
 \label{phase-ii-ADMM-inexact-vnew}
 \begin{array}{r}
 \bv_n^{\ell+1} = {\rm argmin} ~ \langle  {\boldsymbol\xi}_n^{\ell},   \bv_n\rangle   +\frac{\eta}{2}\| \bv_n- \bv_n^{\ell}\|^2 
 =  \bv_n^{\ell} -  \frac{1}{\eta} {\boldsymbol\xi}_n^{\ell},
 \end{array}
\end{eqnarray}
and   update $ \bu_i ^{\ell+1}$ by
 \begin{align}
\eqspace{1.5}\label{phase-ii-ADMM-inexact-un}
 \begin{array}{lll}
 \bu_i ^{\ell+1} &=&{\rm argmin} ~ \langle  {\boldsymbol \zeta}_i ^{\ell}, \bu_i \rangle + 
\frac{\gamma}{2}\|\bu_i -\bu^{\ell+1}\|^2  \\
&+&  \langle \bta_i ^{\ell}, \bu_i \rangle + 
\frac{\mu}{2}\|\bu_i -\bu^{\ell+1}\|^2 \\
 &=&  \bu^{\ell+1}  -\frac{1}{\gamma+\mu}(    {\boldsymbol \zeta}_i ^{\ell} +  \bta_i ^{\ell}).
 \end{array}
\end{align}
Overall, the algorithmic framework to solve problem (\ref{phase-ii-opt}) is presented in Algorithm \ref{algorithm-ADMM-II}.
\begin{figure}[h!]
\centering
\includegraphics[width=0.3\textwidth]{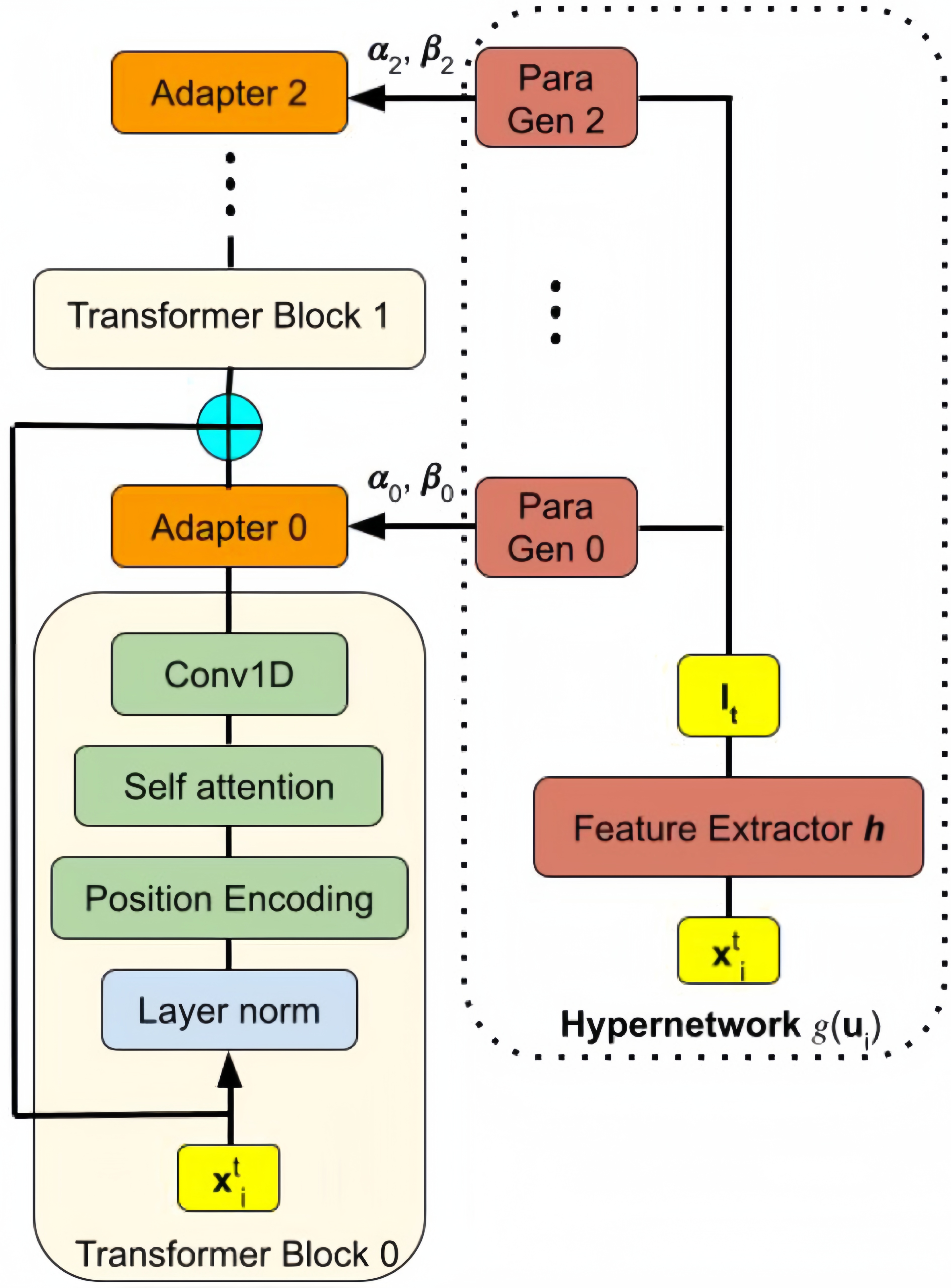}
\caption{Left: The base model architecture and its input received signal $\bx_i^t$. Right: The hypernetwork architecture. `Para Gen $k$' ($k=0,1,2$) represents the parameter generator network, with its generated scale vectors $\boldsymbol{\alpha}_k$ and shift vectors $\boldsymbol{\beta}_k$ for the $k$th adapter layer. The input of hypernetwork $g(\cdot)$ is another received signal $\bold{x}^t_i$. Signals $\bold{x}^t_i$ for both transformer block and hypernetwork are received from the $i$th environment, ${i\in\N}$. }
\label{fig:base_model}\vspace{-3mm}
\end{figure}
\begin{algorithm}[!th]
\SetAlgoLined
{\noindent \justifying Initialize $(\U^0, \T^0)$, $\eta, \gamma$, and $\mu>0$. Set $\ell = 0$.   }

\For{$\ell=0,1,2,\cdots $}{

Update  $\bu^{\ell+1}$ by the first equation in \eqref{phase-ii-ADMM} .\\

Update $\bv_n^{\ell+1}$  by \eqref{phase-ii-ADMM-inexact-vnew}.\\

\For{$i\in\N $}{ 
Update $\bu_i ^{\ell+1} $ by \eqref{phase-ii-ADMM-inexact-un}.\\
Update $\bta_i ^{\ell+1} $  by the last equation in  \eqref{phase-ii-ADMM} .\\
 } 
} 
Return $\bu^{\ell}$ and $\bv_n^{\ell} $.
\caption{iADMM  solving problem (\ref{phase-ii-opt}). \label{algorithm-ADMM-II}}
\end{algorithm}

\subsection{Experiment}\label{OFDM Receiver Result}
\subsubsection{System settings} 
\begin{figure}[th]
\centering
\includegraphics[width=0.475\textwidth]{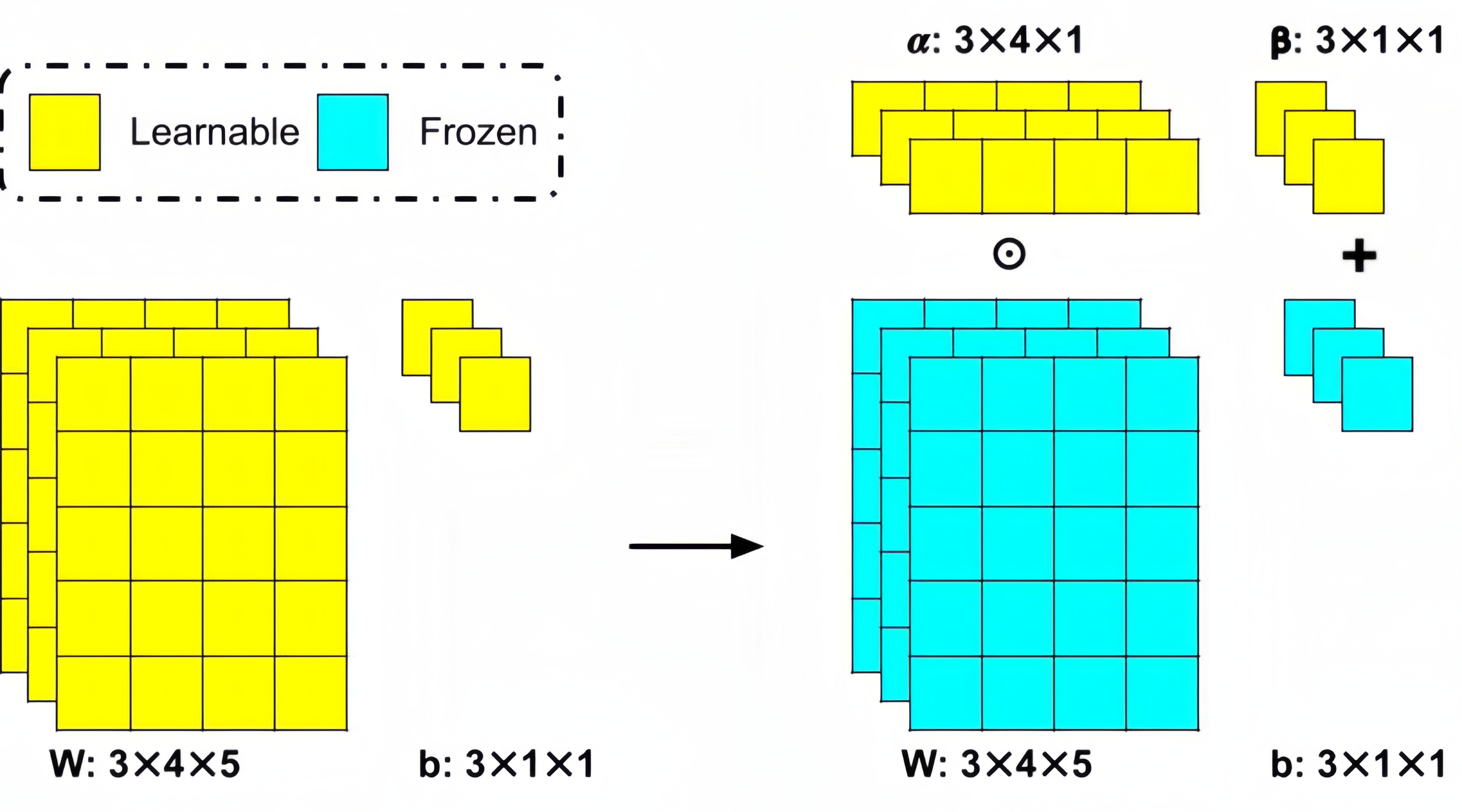}
\caption{Particular parameters $\bold{v}$ fine-tuning process. Here $\bold{W}$ and $\bold{b}$ represent adapter layer's weights and bias, and  $(\boldsymbol{\alpha},\boldsymbol{\beta})$ stands for particular parameter $\bv$. Left: In the conventional TL fine-tuning approach, all parameters are updated. Right: Our channel-level scaling and shifting operations reduce the number of learning parameters and avoid over-fitting.}
\label{fig:scale_shift}
\end{figure}

We apply the framework of EA to the application in the OFDM receiver. In this scenario, we assume that pilot symbols are placed in the first block of a frame, with subsequent blocks containing data. While the channel remains constant within a frame, variations can occur between frames. The OFDM receiver processes one pilot block and one data block in each frame to recover transmitted data in an end-to-end fashion. The OFDM receiver is constructed upon the advanced T5 transformer model \cite{raffel2020exploring}, with the received signal from the $i$th environment with input $\bx_i$ and the transmitted signal as output $\by_i$. The structure of the OFDM receiver with the hyper-network is demonstrated in Fig. \ref{fig:base_model} and detailed design philosophy is same as that in \cite{wang2023efficient}. All transformer blocks share the same parameters across all environments, referring to shared parameter $\bw$. Similar to the approach introduced in \cite{sun2019meta}, we learn to scale $\boldsymbol{\alpha}_k$ and shift $\boldsymbol{\beta}_k$ for each set of weights and biases to fine-tune the $k$th adapter layer, as shown in Fig. \ref{fig:scale_shift}. These fine-tuning parameters ${\{(\boldsymbol{\alpha}_k, \boldsymbol{\beta}_k):k=0,1,2\}}$ refer to particular parameters $\bv$. Unlike conventional fine-tuning methods updating all parameters of the adapter layers, our approach updates only one parameter per channel. 
For example, for a 5-channel convolutional layer, the scale parameter that needs to be learned can be set to the original one-fifth. The parameter vector for fine-tuning convolutional weights is represented as scale $\boldsymbol{\alpha}_k$, initialized to 1 and the bias parameter as shift $\boldsymbol{\beta}_k$, initialized to 0. The final output is obtained by multiplying the adapter weights with $\boldsymbol{\alpha}_k$, together with adding $\boldsymbol{\beta}_k$ to the adapter bias.  

We consider the symbol-spaced multipath channel for the OFDM receiver case. The mean power associated with each multipath component relies on propagation delay $\tau_l$, characterized by a power delay profile (PDP) $P(\tau_l)$ \cite{rappaport1996wireless}. which captures statistics related to small-scale multipath channels. We define the propagation environment as an area shared with the same PDP and simulate multiple environments by changing $\{P(\tau_l):l=1,2,\cdots,L\}$. Consequently, the received signal can be represented as
\begin{eqnarray}\label{equation:channel}
\begin{array}{l}
  r(t) = \sum_{l=1}^L \sqrt{P(\tau_l)}\alpha_l(t)e^{-j\vartheta_l(t)}s(t-\tau_l)+\epsilon(t),\end{array}
\end{eqnarray}
where ${s(t)}$ represents the transmitted signal at time $t$, $L$ denotes the number of paths, $\epsilon(t)$ stands for additive white Gaussian noise, and $\alpha_l(t)$, with $E|\alpha_l(t)|^2 = 1$, characterizes the time-varying variation in path attenuation.

\subsubsection{Experiment configurations} 
\begin{figure}[th]
\centering
\includegraphics[width=0.45\textwidth]{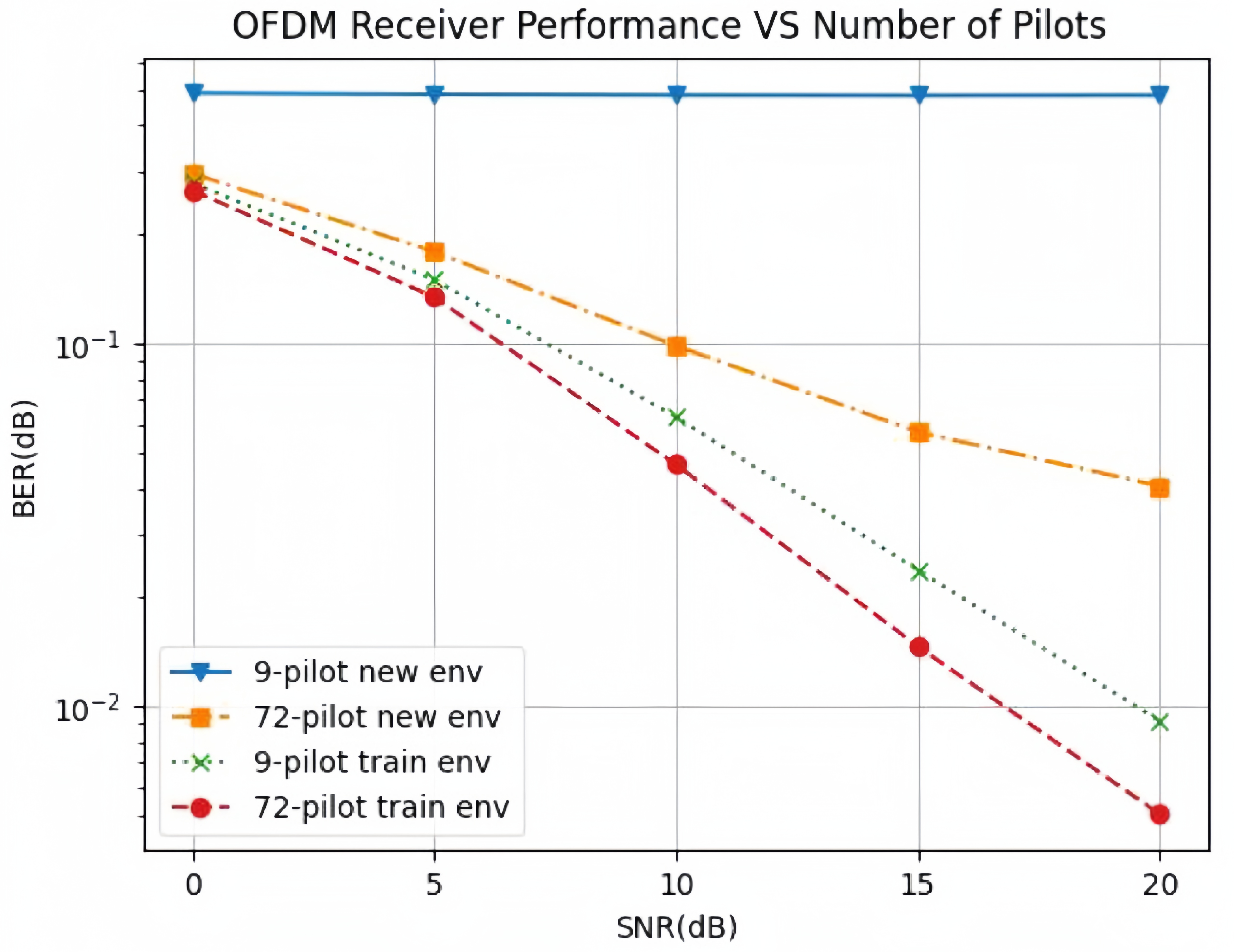}
\caption{The OFDM receiver testing performance with different environments and a number of pilot symbols.}
\label{fig:pilot}
\end{figure}

For this experiment, we consider the OFDM system with 72 sub-carriers and 9 pilot symbols per frame. Fig. \ref{fig:pilot} illustrates that the receiver performs well in previous environments. When tested in new environments, the receiver provides decent performance with the 72-pilot frame while it fails to accurately capture channel features using only 9 pilots. Therefore we choose 9 pilots per frame to demonstrate the superiority of our FSL approach. We follow the wireless world initiative for a new radio model \cite{meinila2009winner} for the wireless channel, using an indoor propagation scenario with a 300m $\times$ 300m map. We consider a single-input-single-output system, in which the positions of the base station and user are varied, along with changes in propagation conditions, resulting in different environments. The number of training environments is ${n=60}$ and each PDP contains ${d_i=500},{i\in\N:=\{0,1,\cdots,59\}}$ 
different instantaneous channel coefficients. We set the number of few-shot samples as ${d_n=16}$ in this stage. The parameters of each model and hyper-parameters of the algorithm are listed in Table \ref{table:2}. Loss function $\ell$ for phases EA-I and EA-II is the binary cross entropy. We test the application of the OFDM receiver where two new environments (i.e., new environments 0 and 1) are encountered.   Moreover, new environment 1 is generated to be more similar to the training environments compared to new environment 0.

\begin{table}[h]
    \centering
    \caption{Model Parameters}\label{table:2}
    \begin{tabular}{c|c}
        \hline
        Parameters & Values\\
        \hline
       Transformer Blocks Conv1D & 128, 128, 128, 128, 128  \\ 
       Learning Rate $\sigma$, $\rho$  & 25, 25\\
       Adapter Layers Conv1D & 128, 128, 128 \\
       Hypernetwork Conv1D  & 128, 128, 64, 32\\
       Parameter Generator Dense  & 3$\times$128, 3$\times$128, 3$\times$128\\
       Learning Rate $\eta$, $\gamma$, $\mu$ & 50, 50, 50 \\
       
       \hline
    \end{tabular} 
    \end{table}

\subsubsection{Adaptation performance and comparison}  
\begin{figure}[th]
\centering
\includegraphics[width=0.45\textwidth]{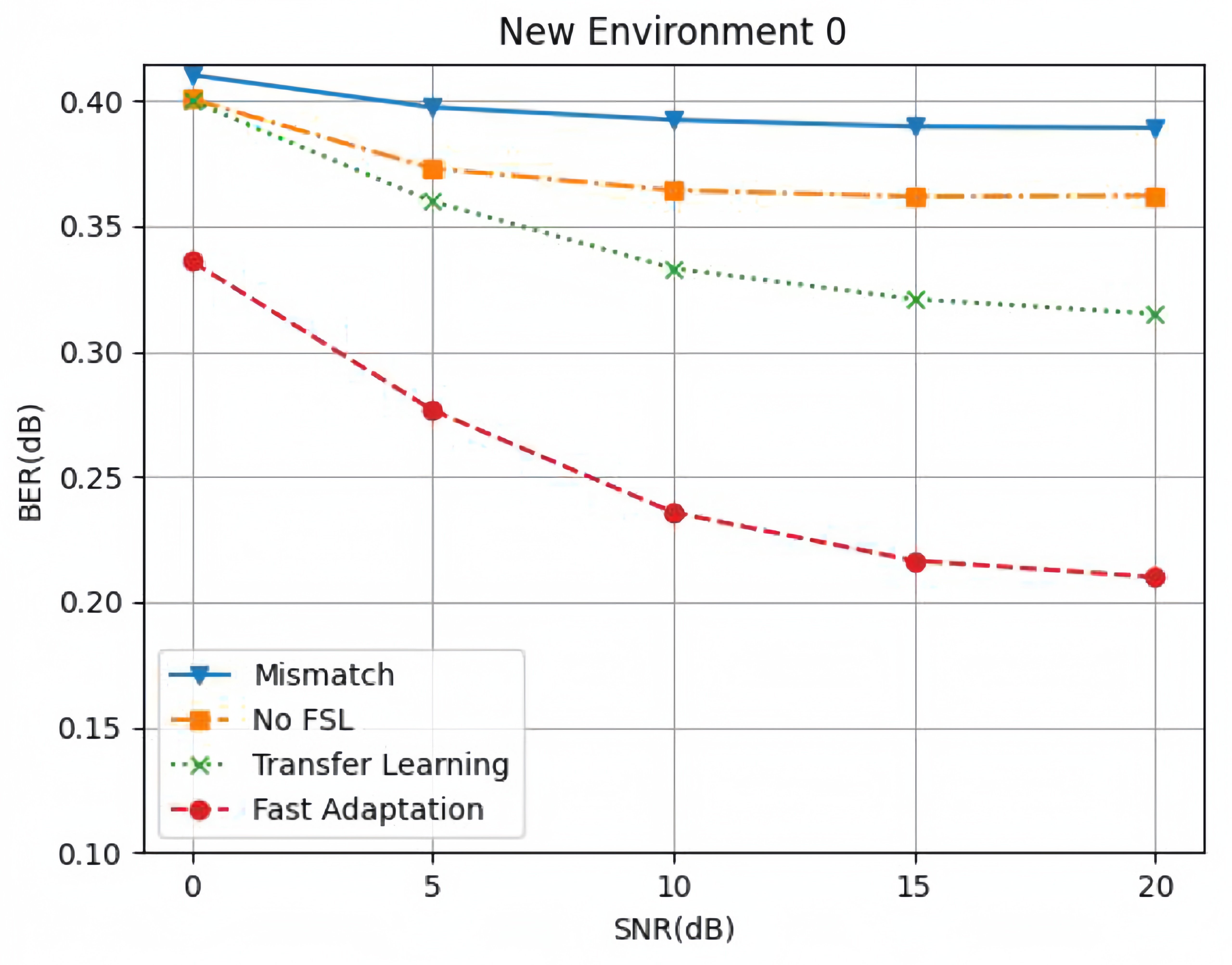} 
\includegraphics[width=0.45\textwidth]{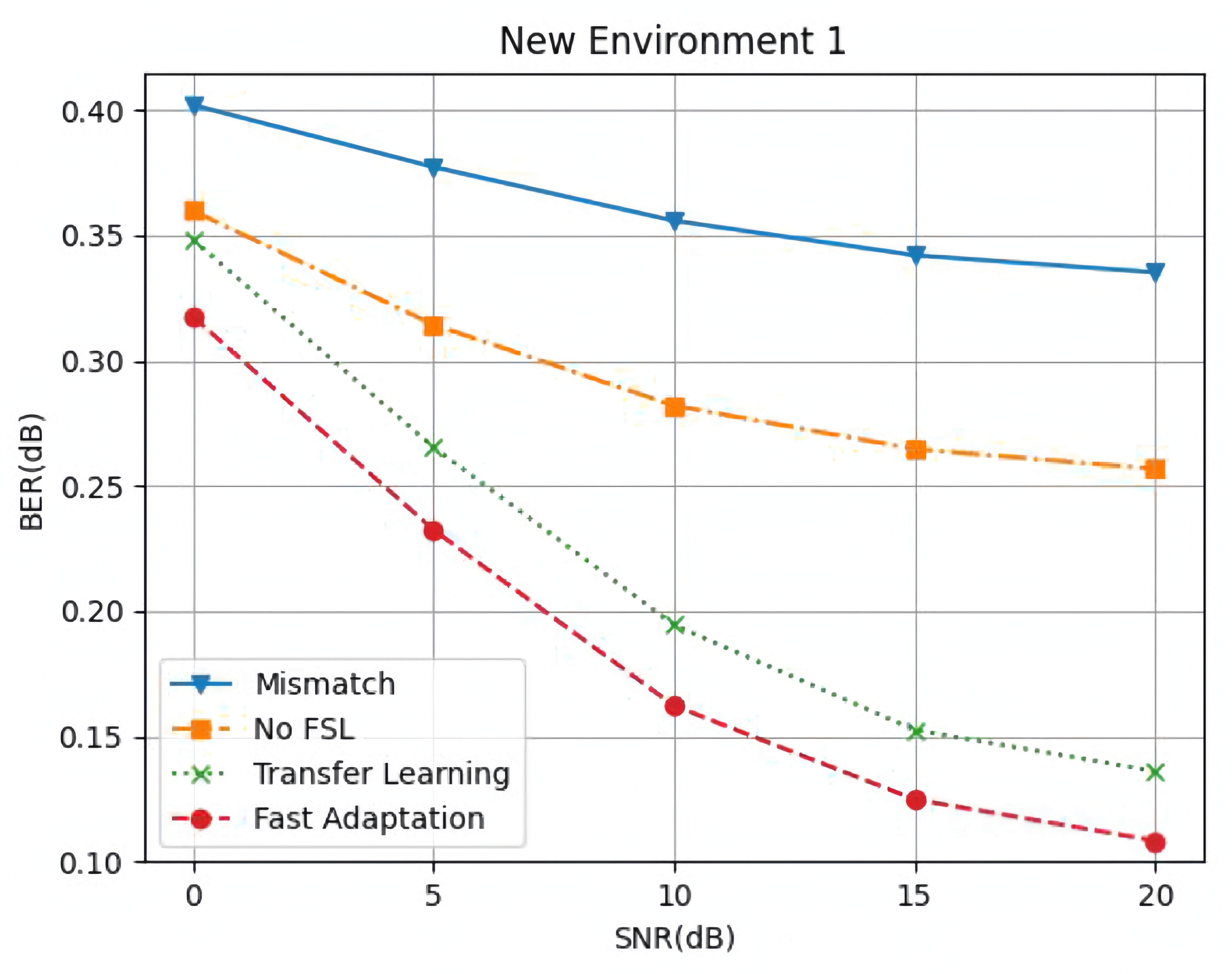}
\caption{Performance of different approaches for the OFDM receiver with two new environments}\label{fig:new_env}
\end{figure}

The mismatch experiment involves training the OFDM receiver in the environment from the training dataset and deploying it directly into new environments. As shown in Fig. \ref{fig:new_env}, the OFDM receiver performs poorly in both new environments, implying significant difference in environment features between the training and testing environments. The OFDM receiver fails to capture the new environment features from pilot blocks due to the few pilot symbols contained in each frame.

Instead of having particular parameters $\bold{v}_i$ for each individual environment ${i\in\N}$, we retrain the entire OFDM receiver for all 60 environments, aiming to capture similarity between the new environments and all 60 training environments. We then deploy this more generalized OFDM receiver into new environments without using any few-shot samples. The testing performance, labeled ``No FSL", verifies that new environment 1 has more similarities to the training environments compared to new environment 0.

We also compare our method with the conventional TL approach, which uses few-shot samples from the new environment to fine-tune all particular parameters ${\bold{v}_i,i\in\N}$ while keeping the remaining ones. TL can improve performance when the training and testing environments have high similarities. This is because the receiver can learn common environment features shared among them during pre-training. However, when the training and testing environments differ significantly (e.g., 60 training environments v.s. new environment 0), TL may not work well as the model may not be able to capture relevant features and patterns from the new environment with few-shot samples. Therefore, the TL approach shows great improvement for new environment 1. 

It can be clearly seen from Fig. \ref{fig:new_env} that our adaptation approach outperforms the TL method. Particularly, the performance of our approach is more pronounced in scenarios where dissimilarities between the training and testing environments are more substantial. In equation \eqref{phase-ii-opt}, there are two parts generating new environment particular parameters $\bold{v}_n$. The first part, $f^\varphi_n(\bu,\bv_n)$, aims to preserve consistency among the new environment data by making the environment feature embedding from each few-shot sample similar. The second part, $f^\phi_n(\bw^*,\bv_n)$, evaluates $\bold{v}_n$ for its generalization ability in the new environment using few-shot samples, which is consistent with the TL fine-tuning approach. We emphasize that $\bold{v}_n$ enables desirable performance of the training model in new environments since the hypernetwork acts as a meta-learner that learns to make the OFDM receiver generalize to various environments. 
\begin{figure}[th]
\centering
\includegraphics[width=0.475\textwidth]{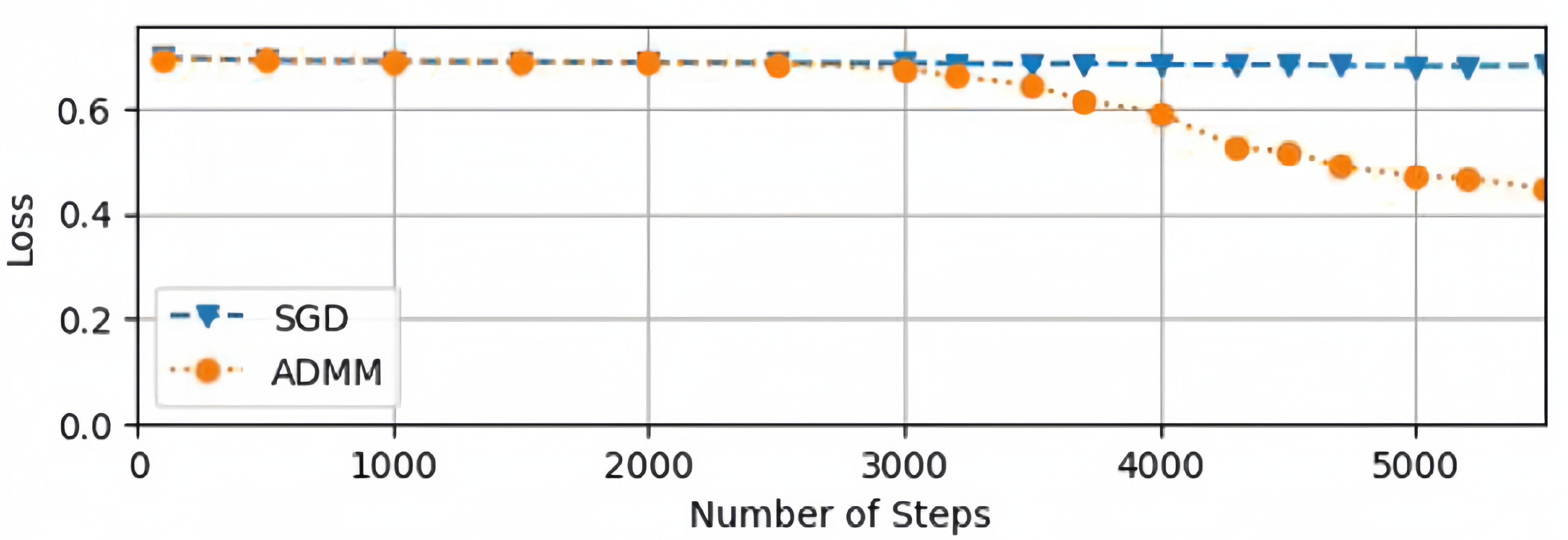}\\\vspace{2mm}
\includegraphics[width=0.475\textwidth]{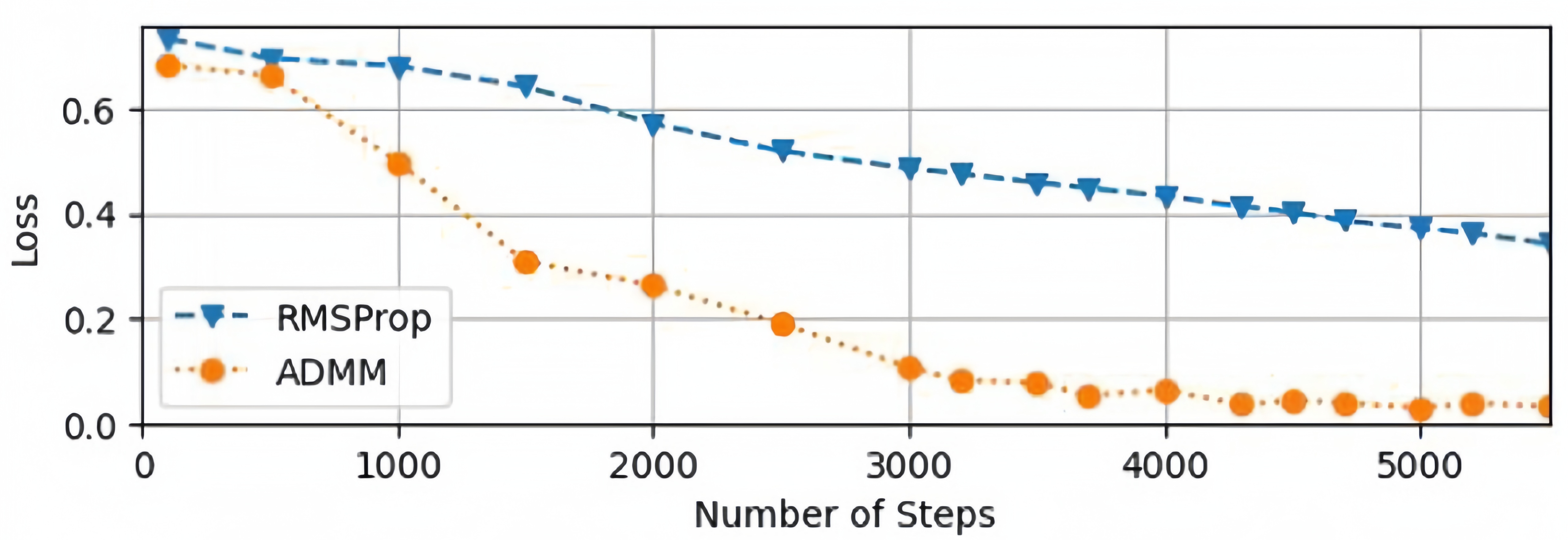}
\caption{Training loss for iADMM and conventional DL optimizers over the first 5500 steps.}\label{fig:loss_curve}
\end{figure}
It can capture the environment features based on previous experience and provide high adaptability for new environments, thereby delivering greater improvement when training and testing environments are less similar.
\subsubsection{Some extension}  

We conduct a comparison between Algorithm \ref{algorithm-ADMM} and two conventional DL optimizers, SGD and RMSprop in our OFDM Receiver application scenario. We employ SGD without incorporating the second-order moment of gradients in training the model. Furthermore, we employ RMSprop, a GD-based approach benefiting from the second-order moment. Regarding Algorithm \ref{algorithm-ADMM}, it can be readily extended by incorporating the second-order moment. To implement this extension, we make modification to \eqref{phase-i-ADMM} and \eqref{phase-i-ADMM-inexact} as follows:
\begin{eqnarray}\label{second-moment}
\eqspace{1.5}
\begin{array}{lll}
\textbf{r}^{\ell+1} &= \varpi\textbf{r}^{\ell} + (1-\varpi)\textbf{g}^{\ell}\odot\textbf{g}^{\ell}\\
\bw^{\ell+1} &=  \frac{1}{n}\sum_{i=0}^{n-1}\Big(\bw_i^{\ell}+\frac{1}{\sigma(\sqrt{\textbf{r}^{\ell+1}}+\epsilon)}\odot\bpi_i^{\ell}\Big), \\
\bw_i^{\ell+1} &= \bw^{\ell+1}-\frac{1}{(\rho  + \sigma)(\sqrt{\textbf{r}^{\ell+1}} + \epsilon)  } \odot ( {\boldsymbol  \zeta}_i ^{\ell}+\bpi_i ^{\ell} ),~i\in\N,\\
\bv_i ^{\ell+1}&= \bv_i ^{\ell}   -\frac{1}{\rho }   {\boldsymbol  \xi}_i ^{\ell},~i\in\N,\\
\bpi_i ^{\ell+1} &= \bpi_i ^{\ell} + \sigma (\bw_i ^{\ell+1}-\bw^{\ell+1}),~i\in\N,
\end{array}
\end{eqnarray}
where ${\textbf{g}^{\ell}:=\frac{1}{n} \sum_{i=0}^{n-1} \nabla_{\bw}f^\phi_i (\bw^\ell ,\bv_i^\ell )}$ and thus $\textbf{r}^{\ell}$ refers to the cumulative squared gradient with ${\textbf{r}^0=0}$, and ${\varpi\in(0,1)}$ denotes the attenuation coefficient, controlling the retention of historical information during parameter updates. Here $\odot$ is the Hardamard product and ${1/\sqrt{\textbf{r}}:=(1/\sqrt{r_1},\cdots,1/\sqrt{r_m})^\top}$ for ${{\bf r}\in\R^m}$.  We have plotted the training loss for the first 5500 steps, as depicted in Fig. \ref{fig:loss_curve}, showcasing the averaged training loss across all environments. In the figure above,  iADMM without using the second-order moment (i.e., Algorithm \ref{algorithm-ADMM}) achieves quicker convergence than SGD. When incorporating the second-order moment into iADMM (according to updates \eqref{second-moment}), it outperforms RMSProp, as shown in Fig. \ref{fig:loss_curve}.

\section{Online Adaptation (OA) with Application in BF Prediction}\label{OA:BF}
Similar to Section \ref{EA_OFDM}, this section introduces the OA learning scheme and algorithm. The experiment for the OA application to the BF prediction is also included. 
\begin{figure}[h]
\centering
\includegraphics[scale=.395]{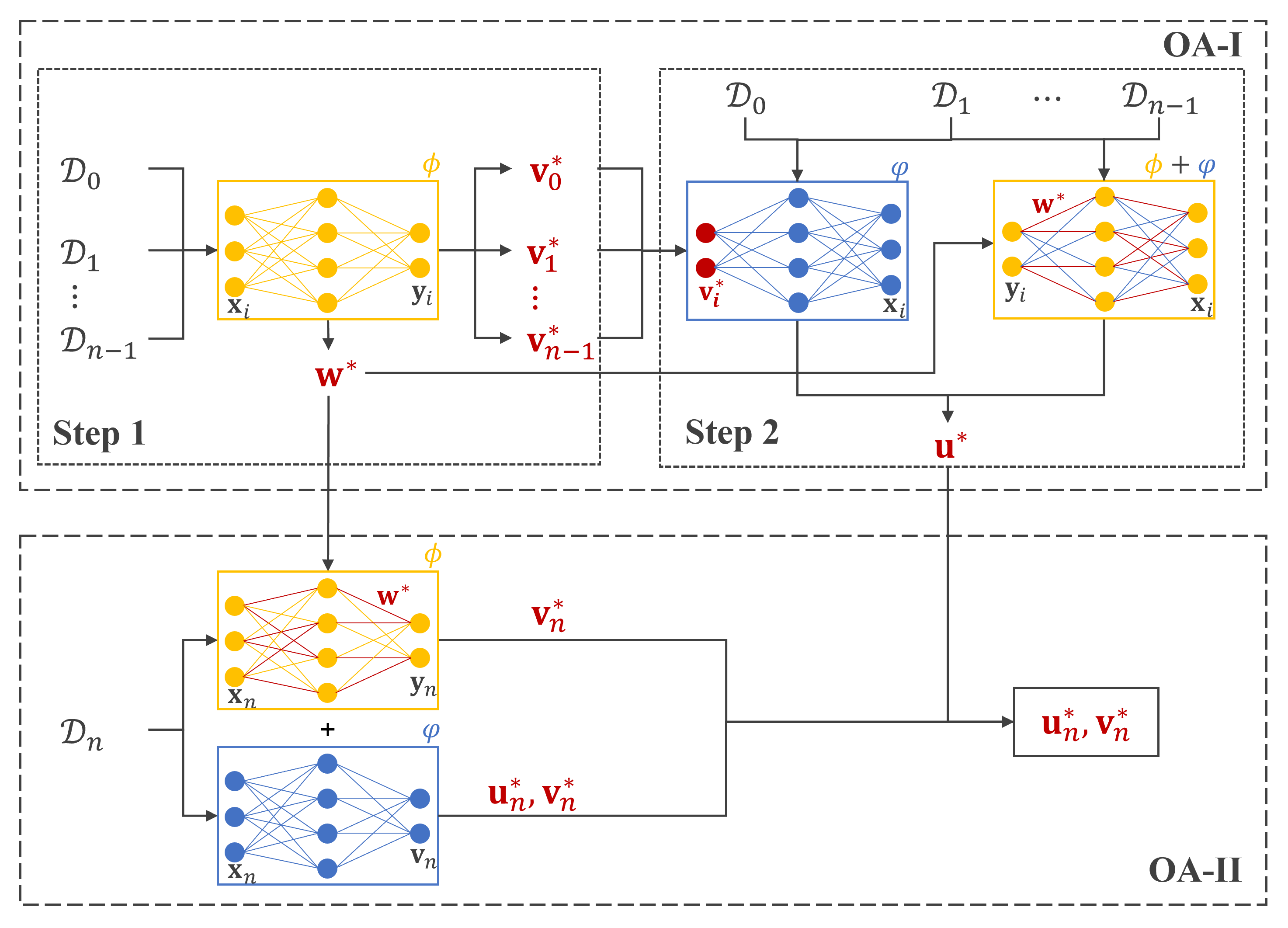}
\caption{Structure of OA.}\label{fig:OA}
\end{figure}
\subsection{Problem Formulation}
This approach also consists of two phases. The first phase includes two steps and the second phase adapts the information to the new environment only using its few-shot samples, as shown in Fig. \ref{fig:OA}.  

{\bf OA-I: Learning from previous environments} At first, similar to EA-I, we learn a shared parameter $\bw^*$ and $n$ particular parameters $( {\bv}_0^*,{\bv}_1^*,\cdots, {\bv}_{n-1}^*)$ by addressing problem \eqref{phase-i-opt}. Then we learn a shared parameter $\bu^*$ offline to characterize the relationship as in \eqref{relationship-hypernetwork} between ${\bv}_i ^*$ and $\bx_i $ for every group $n$. 
Similar to \eqref{phase-ii-opt}, this can be done by solving
\begin{eqnarray}\label{phase-ii-opt-1}
\eqspace{1.5}
\begin{array}{r}
{\bu}^*= \underset{\bu}{\rm argmin}~\frac{\lambda }{n}\sum_{i=0}^{n-1}\sum_{t=1}^{d_i } \ell\left(\phi(\bw^*,\varphi(\bu;\bx_i ^{t}) ;\bx^{t}_i ),\by^{t}_i  \right) \\
+\frac{1}{n}\sum_{i=0}^{n-1} f^\varphi_i (\bu,\bv_i ^*),
\end{array}
\end{eqnarray}
where $f^\varphi_i $ is defined as \eqref{obj-phi-new}. This problem minimizes the gap between  ${\bv}_i ^*$ and  $\varphi(\bu;\bx_i ^{t})$ while maintaining the training loss when ${\bv}_i $ is replaced by  $\varphi(\bu;\bx_i ^{t})$ in the objective function of \eqref{phase-i-opt}. It is worth pointing out that differing from \eqref{phase-ii-opt}, the above problem does not involve new environmental data, so it is still an offline training process.

{\bf OA-II: Adaptation into the new environment} When facing a new environment, we expect a relationship between ${\bv}_n^*$ and $\varphi$ as \eqref{relationship-hypernetwork} and a small training loss $f^\phi_n(\bw^*,\bv_n)$. Therefore, we focus on the following optimization problem,
\begin{align}\label{phase-iii-opt}
\eqspace{1.5}
\begin{array}{r}
({\bu}^*_n,{\bv}^*_n):=\underset{\bu_n,\bv_n}{\rm argmin} ~   c_1f^\varphi_n(\bu_n,\bv_n)+f^\phi_n(\bw^*,\bv_n)\\
+ c_2\left\|{\bu_n}-{\bu}^*\right\|^2 +  c_3\left\|{\bu_n}\right\|^2+ c_4\left\|{\bv_n}\right\|^2,
\end{array}
\end{align}
where  ${ c_i>0}$,for ${i=1,2,3,}$ and ${4}$ are four penalty parameters, $f^\phi_n$ and $f^\varphi_n$ are defined in \eqref{obj-phi-new}. The objective function in \eqref{phase-iii-opt} consists of four components.  The purpose of using the first two components is similar to that of \eqref{phase-ii-opt-1}. Third component $\|{\bu_n}-{\bu}^*\|^2$ aims to leverage prior information from ${\bu_n}^*$,   $\|{\bu_n}\|^2$ and $\|{\bv_n}\|^2$ serve as $L_2$ regularization to prevent overfitting based on limited $d_n$ samples. It is evident that this phase relies solely on the few-shot samples, i.e., $\D_n$, and is built upon shared parameter $\bw^*$ and prior knowledge ${\bu}^*$ because the new environment exhibits similar patterns to the previous $n$ groups of environments.

Comparing Fig. \ref{fig:EA} and \ref{fig:OA}, we can see the difference between EA and OA frameworks. Both steps in phase OA-I belong to the offline training using  $N$ known environmental datasets. In phase OA-II, we fully leverage prior knowledge $\bw^*$ and $\bu^*$ to learn a particular parameter $\bv_n^*$ purely based on the new environmental data. When encountering different new environments, we only need to carry out the training for phase OA-II, thereby saving a significant amount of energy.

\subsection{Solving problem (\ref{phase-ii-opt-1})}
We take advantage of the model-agnostic meta-learning (MAML)  algorithm proposed in \cite{finn2017model} to solve problem (\ref{phase-ii-opt-1}).

\vspace{-2mm}
\subsection{Solving problem (\ref{phase-iii-opt})}
We adopt iADM to solve problem \eqref{phase-iii-opt}. Denote   
\begin{align*} 
\eqspace{1.5}
\begin{array}{lrl}g(\bu_n,\bv_n)&:=& c_1f^\varphi_n(\bu_n,\bv_n)+  c_3\left\|{\bu_n}\right\|^2,\\
{F}(\bu_n,\bv_n)&:=& 
g(\bu_n,\bv_n) + f^\phi_n(\bw^*,\bv_n)\\
&+&  c_2\left\|{\bu_n}-{\bu}^*\right\|^2+ c_4\left\|{\bv_n}\right\|^2.  \end{array}
\end{align*}
Here, ${F}(\bu_n,\bv_n)$ is the objective function of problem \eqref{phase-iii-opt}. Now let ${\tau^{\ell}>0}$ and ${\kappa^{\ell}>0}$ be two chosen constants at the $\ell$th iteration. Let $ {{\boldsymbol\zeta}_n^{\ell}\in \partial_{\bu} g(\bu_n^{\ell},\bv_n^{\ell})} $ and  $ {{\boldsymbol\xi}^{\ell}_n\in \partial_{\bv}     f^\phi_n(\bw^*,\bv_n^{\ell})}$. By initializing ${(\bu^0,\bv^0_n)}$ and for every ${\ell=0,1,2,\cdots}$, iADM updates $\bu^{\ell+1}_n$  by
\begin{align}
\label{phase-iii-ADM-u} 
\bu_n^{\ell+1}
 &={\rm argmin}~\langle  {\boldsymbol\zeta}_n^{\ell}, \bu_n \rangle +  \tau^{\ell} \|{\bu}_n-{\bu}_n^{\ell}\|^2+ c_2 \|{\bu}_n-{\bu}^*\|^2 \nonumber\\[1.25ex]
&= \begin{array}{rrr}\frac{1}{\tau^{\ell}+ c_2} (\tau^{\ell}\bu_n^{\ell} +  c_2\bu^* -\frac{1}{2}  {\boldsymbol\zeta}_n^{\ell} )\end{array},
\end{align} 
and update $\bv^{\ell+1}_n$  by
\begin{eqnarray}
\label{phase-iii-ADM-v} 
\eqspace{1.5}
\begin{array}{rrr}
\bv^{\ell+1}_n
={\rm argmin}~ c_1 f^\varphi_n(\bu^{\ell+1},\bv_n) + \langle {\boldsymbol\xi}^{\ell}_n, \bv_n\rangle\\
+ \kappa^{\ell} \|\bv_n-\bv^{\ell}_n\|^2+  c_4 \|\bv_n\|^2 \\
=\frac{1}{ c_1 d_n+ c_4+\kappa^{\ell}}( c_1\sum_{i=1}^{d_n}  \varphi(\bu^{\ell+1};\bx^{t}_n)+ \kappa^{\ell}\bv^{\ell}_n -\frac{1}{2} {\boldsymbol\xi}^{\ell}_n). 
\end{array}
\end{eqnarray}
 Overall, the algorithmic framework of the above updates is presented in Algorithm \ref{algorithm-ADMM-III}.
\begin{algorithm}[!th]
\SetAlgoLined
{\noindent \justifying Initialize ${\bu_n^0=\bu^*}$, ${\bv^0_n=\frac{1}{d_n}\sum_{t=1}^{d_n}\varphi(\bu^*;\bx_n^{t})}$, and ${ c_i>0,i=1,2,3,4}$. Set $\ell = 0$. }  

\For{$\ell=0,1,2,\cdots $}{

Choose $\tau^{\ell}$ and update  $\bu^{\ell+1}_n$ by \eqref{phase-iii-ADM-u}.

Choose $\kappa^{\ell}$ and update    $\bv_n^{\ell+1}$  by \eqref{phase-iii-ADM-v}.

} 
Return $\bu_n^{\ell}$ and $\bv^{\ell}_n$.
\caption{iADM solving problem \eqref{phase-iii-opt}.  \label{algorithm-ADMM-III}}
\end{algorithm}

The following theorem, proved in Appendix, shows the convergence results of Algorithm \ref{algorithm-ADMM-III} under reasonable conditions.

\begin{figure*}[th]
\centering
\includegraphics[width=0.8\textwidth]{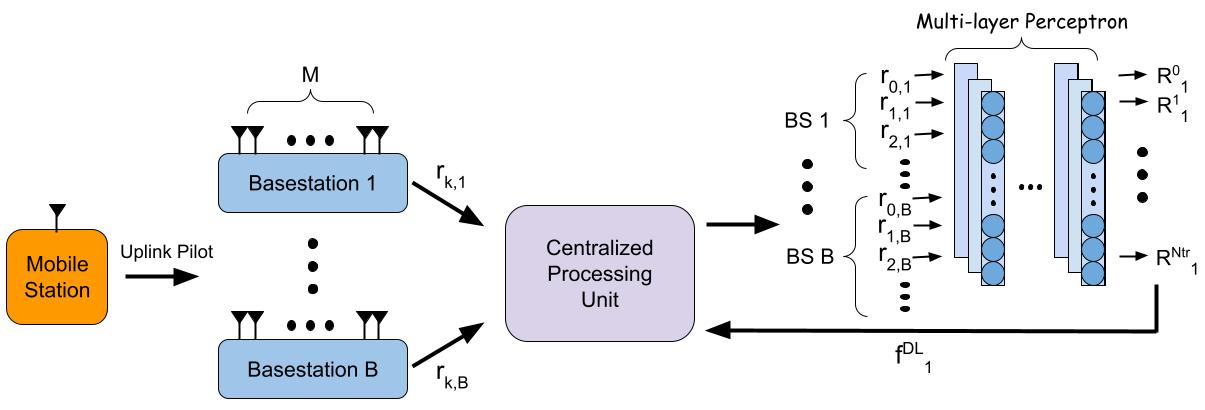}
\caption{The operation of the proposed BF prediction system.  This figure is an example of BS 1 using DL model. The centralized processing unit obtains the received pilot signal from all BSs and asks the DL model to predict $\textbf{f}_b^{DL}$ and feedback. Each BS employs a single RF chain and implements analog-only beamforming through networks of quantized phase shifters.}
\label{fig:BF_system}
\end{figure*} 
\begin{theorem*}\label{the-iADM}
Suppose $f_n^\varphi$ and $f_n^\phi$ are twice continuously differentiable. Choose ${\tau^\ell>\tau^*}$ and ${\kappa^\ell> \kappa^*}$ for any ${\ell\geq 0}$, where $\tau^*$ and $\kappa^*$ are defined as (\ref{def-tau-k}).   Then the sequence of objective function values $\{F(\bu_n^\ell,\bv_n^\ell)\}$ is strictly decreasing and converges. Moreover, $\lim_{\ell\to\infty}(\bu_n^{\ell+1}-\bu_n^\ell)=0$ and $\lim_{\ell\to\infty}(\bv_n^{\ell+1}-\bv_n^\ell)=0$.
\end{theorem*}

It is worth mentioning that there are many neural networks satisfying the twice continuous differentiability in Theorem above. For example, if activation functions (e.g., sigmoid and tanh functions) are twice continuously differentiable used in neural networks $\phi$ and $\varphi$, then $f_n^\varphi$ and $f_n^\phi$ are twice continuously differentiable as well.  

\subsection{Experiment}\label{BF_Prediction}

\subsubsection{System settings}  We apply the scheme of OA to process BF prediction. In the mmWave communication system, $B$ base stations (BSs) with $M$ antennas are concurrently serving a mobile user with a single antenna, and all the BSs are linked to a centralized processing unit. The commercial ray-tracing simulator, DeepMIMO \cite{alkhateeb2019deepmimo}, together with the geometric wideband mmWave channel model \cite{akdeniz2014millimeter}, is employed to simulate propagation environments. There are $L$ clusters to represent a group of closely spaced signal paths. Within each cluster, there is one representative ray selected to capture the characteristics of that cluster. This ray has a time delay $\tau_l$ and azimuth/elevation angles of arrival (AoA) ${\theta_l,\vartheta_l}$, ${l=1,2,\cdots,L}$. Furthermore,  pulse shaping function $p(\tau)$ is employed to evaluate the $T_S$-spaced signaling at $\tau$ seconds. We consider perfect frequency and carrier offset synchronization for users. The user received signal at subcarrier $k$ with transmitted signal $\textbf{s}_{k,b}$ from the $b$th BS and the delay-$d$ channel vector $\boldsymbol{h}_{k,b}$ at $k$th subcarrier can be expressed as
\begin{eqnarray}\label{mmwave_channel}
\eqspace{1.5}
\begin{array}{lll}
r_k &=& \sum_{b=1}^B \langle\boldsymbol{h}_{k,b},\boldsymbol{s}_{k,b}\rangle + \varepsilon_k,\\
\boldsymbol{h}_{k,b} &=& \sum_{d=0}^{D-1} \boldsymbol{h}_{d,b}e^{-j\frac{2\pi k}{K}d},\\
\boldsymbol{h}_{d,b} &=& \sqrt{\frac{M} {\rho_b}}\sum_{l=1}^L {\alpha_l}p(dT_S-\tau_l)\boldsymbol{a}_b(\theta_l, \vartheta_l)
\end{array}
\end{eqnarray}
where $\varepsilon_k$ is the receive noise at $k$th subcarrier, $\rho_b$ denotes the path loss between the user and the $b$th BS,  $\alpha_l$ characterizes the path attenuation, and $\boldsymbol{a}_b(\theta_l, \vartheta_l)$ is the array response vector of the $b$th BS at the AoA $\theta_l, \vartheta_l$. Assuming this mmWave channel is a block-fading channel, where ${\{\boldsymbol{h}_{k,b}:k=1,2,\cdots,K\}}$ stay constants over the channel coherence time.

The BF design strategy and simulation scenario keep aligned with the work proposed in \cite{alkhateeb2019deepmimo}, where each BS ${b\in {\mathbb B}:=\{1,2,\cdots,B\}}$ uses a time-domain analog beamforming ${\textbf{f}^{RF}_b \in \mathbb{C}^{M}}$ to precode the transmitted signal, where `RF' is an abbreviation for radio frequency. The BF vectors, ${\{\textbf{f}^{RF}_b: b\in {\mathbb B}\}}$, are selected from finite-size codebooks $\boldsymbol{\mathcal{F}}_{RF}$ with a size as $N_{tr}$. The main challenge of designing this BF prediction system is to select the BF vector that can maximize the achievable data rate in the high-mobile scenario. We consider the impact of the time overhead required for channel estimation and BF design. Let $T_B$ refer to the beam coherence time. The first $T_{tr}$ seconds is allocated for channel estimation and BF design and the rest of the time in $T_B$ seconds is used for data transmission with designed BF vectors. The detailed proof and analytical steps for formulating the BF design strategy are shown in \cite{alkhateeb2018deep}. Here we directly present solutions for different strategies. A baseline solution based on conventional communication system tools is proposed in \cite{hur2013millimeter}, which directly trains the BF vectors through an exhaustive search to find the best beams. All BF codewords are combined with the repeated pilot sequence sent by users. Then these combined signals are fed back to the central processor, which computes the received power using every RF  BF vector and determines the downlink BF vector $\textbf{f}_b^{BL}$ separately for every BS ${b\in {\mathbb B}}$. Given the individual beam training pilot sequence time as $T_p$ and the overall beam training time $T_{tr}=N_{tr}\times T_p$, the achievable rate can be defined as     
\begin{eqnarray}\label{baseline_rate}
\eqspace{1.5}
\begin{array}{ll}
R^{BL} =& \big(1-\frac{T_{tr}}{T_B}\big)\times\\
&\frac{1}{K}\sum^K_{k=1} \log_2\big(1+\varrho\big(\sum_{b=1}^B |\langle \textbf{h}_{k,b}, \textbf{f}^{BL}_b\rangle |^2\big)^2\big),
\end{array}
\end{eqnarray}
where $\varrho$ is the signal-to-noise ratio.  This achievable rate will be used as the baseline rate in the sequel.

\subsubsection{Experiment configurations} Our BF prediction system is constructed with a DNN. System input $\bx_i$ for each environment ${i\in\N}$ is uplink received pilot sequence $r_{k,b}$ collected from $B$ BSs and $K$ subcarriers. It has been demonstrated in \cite{alkhateeb2018deep} that splitting the BS RF design leads to efficient, low-complexity systems that can achieve rates close to the optimal bound. Therefore, $B$ independent DNNs are implemented for the $B$ BSs. Each DNN has $N_{tr}$ outputs, which represent the predicted achievable rate with all $N_{tr}$ BF vectors. The DNN consists of 8 fully connected layers each with 256 nodes. These layers' weights are considered as shared parameter $\textbf{w}$ and adapter layers are introduced as particular parameters $\textbf{v}$. Particular parameters are integrated between fully connected layers separately and are generated by the hypernetwork. The adapter layer and hypernetwork follow the same mechanism as proposed in Section \ref{OFDM Receiver Result}, where the output of adapter layers are modified by extra weights and bias generated by the hyper-network. This DL coordinated approach contains two uplink processes in the beam training time: the uplink pilot sequence with omni beam pattern and predicted beam $\textbf{f}^{DL}_b$. Therefore, the DL coordinated effective achievable rate can be denoted as
\begin{eqnarray*}
\eqspace{1.5}
\begin{array}{lll}
R^{DL} =& \big(1-\frac{2T_p}{T_B}\big)\times\\
&\frac{1}{K}\sum^K_{k=1}\log_2\big(1+\varrho\big(\sum_{b=1}^B|\langle\textbf{h}_{k,b},\textbf{f}^{DL}_b\rangle|^2\big)^2\big).
\end{array}
\end{eqnarray*}

\begin{figure}[th]
\centering
\includegraphics[width=0.4\textwidth]{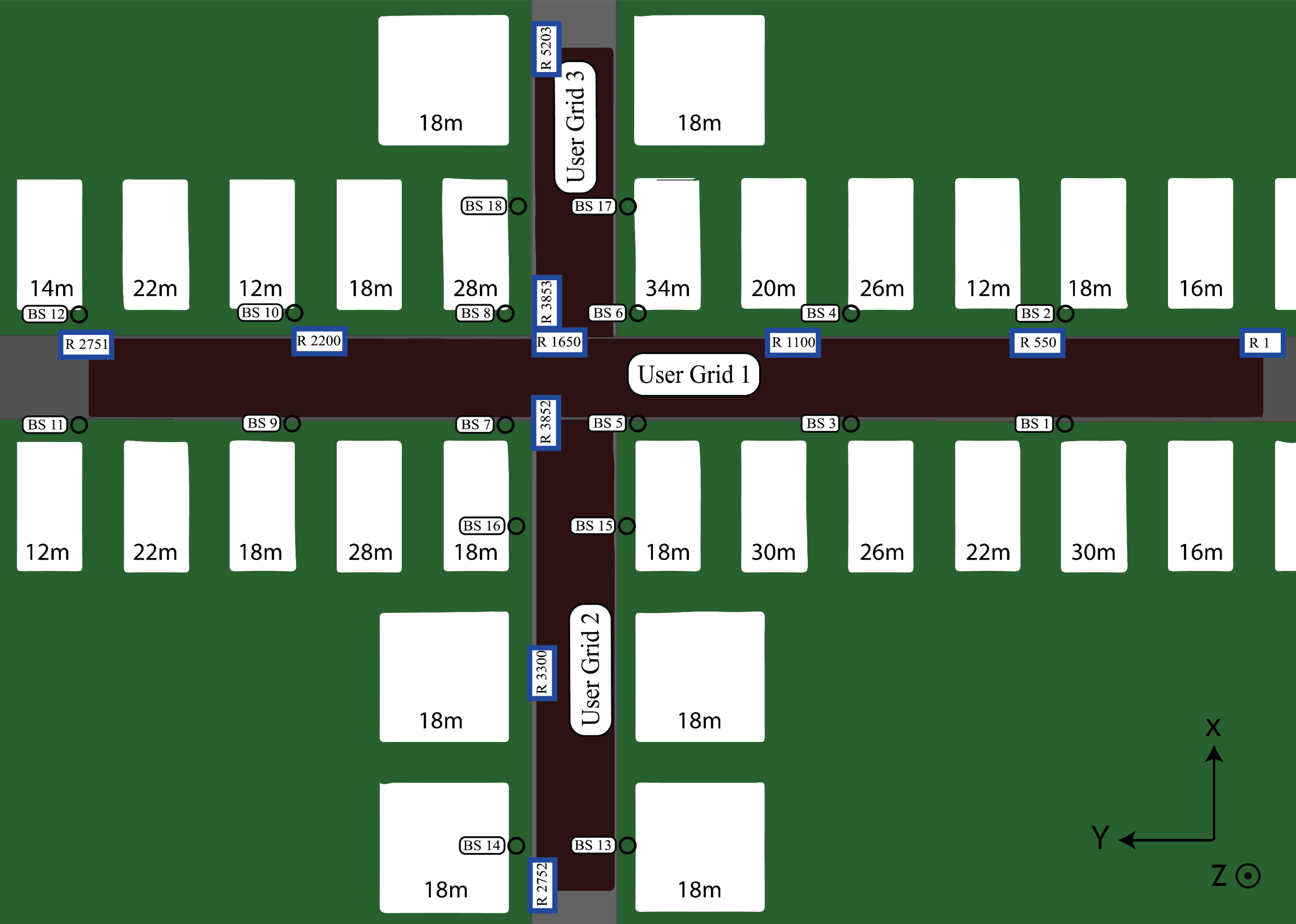}
\caption{The top view of the simulation layout}
\label{fig:O1_topV}
\end{figure}

As depicted in Fig.~\ref{fig:O1_topV}, the simulation operates in a street-level environment. In total, there are $B=18$ different BSs and three different user grids with 1,184,923 users in this simulation scenario. We deploy two BSs to serve one vehicular mobile user simultaneously over the 60 GHz band. Varied propagation environments are generated by selecting different combinations of BSs and users. Notice that the BS pairs chosen for testing are not included within the training dataset. The training dataset encompasses ${n=60}$ different environments, each involving $7240$ mobile users (namely, ${d_i=7204}$ for every ${i\in\N}$). Subsequently, we have two testing environments configured as follows: new environment 0 involves BSs 17 and 18 alongside User Grid 3, while new environment 1 encompasses BSs 14 and 16 paired with User Grid 2.

\subsubsection{Adaptation performance and comparison} 
\begin{figure}[th]
\centering
\includegraphics[width=0.45\textwidth]{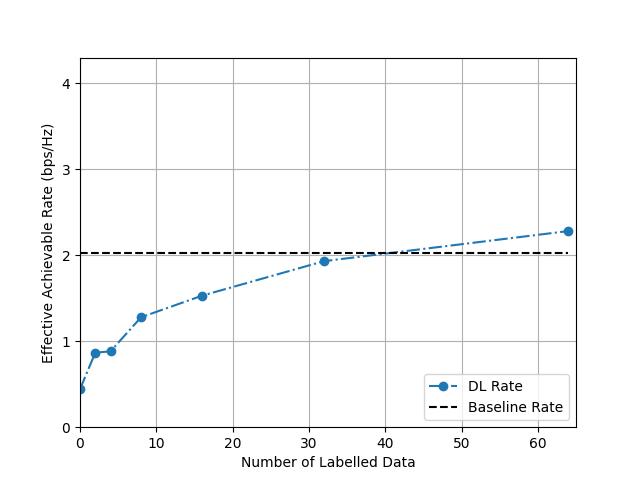}
\caption{The achievable rate versus the number of available samples for the new environment 0.}
\label{fig:fsl_num}
\end{figure}

First, we investigate the impact of the number of available samples on the adaptation to new environments. As depicted in Fig~\ref{fig:fsl_num}, the `Baseline Rate' refers to the baseline coordinated BF using \eqref{baseline_rate}. It is observed that the achievable rate achieved using our proposed algorithm is even lower than the baseline rate when the number of samples falls below 32. Therefore, we choose the number of new environment samples as 64.    

\begin{figure}[th]
\centering
\includegraphics[width=0.45\textwidth]{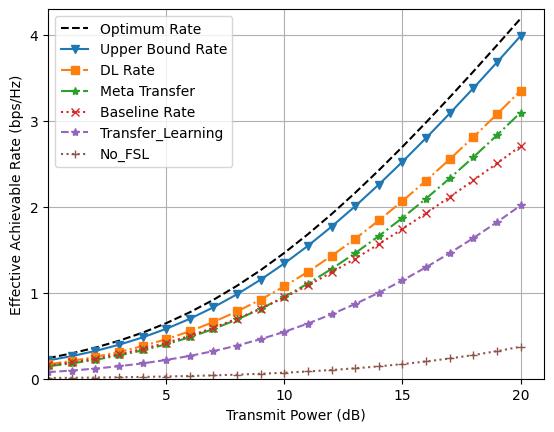}\\
\includegraphics[width=0.45\textwidth]{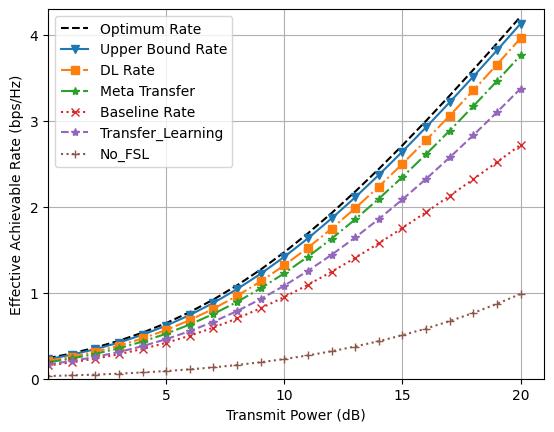}
\caption{The achievable rate generated by different approaches for the BF prediction system in two new environments}\label{fig:mmwave_new_env}
\end{figure}

The performance of our proposed approach is illustrated in Fig.~\ref{fig:mmwave_new_env}, denoted as `DL Rate', and is compared against some state-of-the-art approaches. Similar to the `No FSL' experiment discussed in Section \ref{OFDM Receiver Result}, we assess the similarity between the training and new environments. The testing results indicate that new environment 1 exhibits a considerable degree of similarity to the training set, but the achievable rate remains significantly below the baseline rate. 

TL is implemented utilizing the same mechanism detailed in Section \ref{OFDM Receiver Result}. We can draw the same conclusion that TL achieves better performance if the training and new environments share more similar patterns. Additionally, a more advanced FSL approach is employed called meta-transfer learning (MTL) \cite{sun2019meta}. MTL transfers shared weights $\textbf{w}$ and uses MAML \cite{finn2017model} to learn particular parameters $\textbf{v}_n$. Although this MTL approach achieves better performance than the baseline, it is still worse than our proposed algorithm. This comparison outcome indicates that penalty terms in phase OA-II enhance the generalization ability of the BF prediction systems during the online adaptations.  

In Fig.~\ref{fig:mmwave_new_env}, we establish two upper-bounded achievable rates. `Upper Bound Rate' refers to the achievable rate obtained by the BF prediction system, trained with sufficient data in the new environment, approximately 7,240 samples. `Optimum Rate' demonstrates the data rate calculated according to \ref{baseline_rate} without a time-consuming search for the best BF vector. `Optimum Rate' represents the theoretical best data rate, which is unattainable in practical applications, whereas `Upper Bound Rate' reflects the best achievable rate. The proposed FSL algorithm closely approaches the best achievable rate in the new environment 1. Furthermore, in more dissimilar environments like new environment 0, the proposed algorithm exhibits a substantial improvement compared with other approaches.

\section{Conclusion}\label{sec:conclusion}
We proposed two FSL schemes that enable efficient adaptation to new environments using few-shot samples based on prior knowledge from multiple known environments.  Their efficiency has been evaluated by processing the issues in the context of the OFDM receiver and mmWave BF system. The proposed approaches exhibited superior performance compared to well-established alternatives, particularly in scenarios where the new environments share limited similarity to the known ones. Moreover, the developed frameworks are versatile and might be able to deal with various other FSL tasks in wireless communications.

\bibliographystyle{IEEEtran}
\bibliography{ref}

\section*{Appendix: Proof of Theorem}

Before we give the proof, for a fixed $\bv_n$, let  ${\nabla^2_{\bu\bu} g(\bu_n,\bv_n)}$ be the Hessian matrix of $ {g(\cdot,\bv_n)}$   and $\mathbb{B}(\bu^*,\delta):=\{\bu:\|\bu-\bu^*\|^2\leq \delta\}$. We denote 
\begin{align*} 
\eqspace{1.5}
\begin{array}{lll}
\Omega:=\{(\bu_n,\bv_n):{F}(\bu_n,\bv_n)\leq F^0:={F}(\bu_n^0,\bv_n^0)\},\\
\delta_{\bu}^*:=\frac{3}{ c_2}F^0  + \frac{2}{ c_2^2}\sup_{(\bu_n,\bv_n)\in \Omega}\|\nabla_{\bu} g(\bu_n,\bv_n)\|^2,\\
\delta_{\bv}^*:=\frac{3}{ c_4}F^0  + \frac{2}{ c_4^2}\sup_{(\bu_n,\bv_n)\in \Omega}\|\nabla_{\bv} f^\phi_n(\bw^*,\bv_n)\|^2.
\end{array}
\end{align*}
Based on the above constants, we also define
\begin{eqnarray*}
 \label{def-tau-k}
\eqspace{1.5}
\begin{array}{lll}
\tau^*:=\sup_{(\bu_n,\bv_n)\in \mathbb{B}(\bu^*,\delta_{\bu}^*)\times \mathbb{B}(0,\frac{F^0}{ c_4})}  \frac{1}{2}\|\nabla^2_{\bu\bu} g(\bu_n,\bv_n)\|,\\ 
\kappa^*:=\sup_{\bv_n\in \mathbb{B}(0,\delta_{\bv}^*)} \frac{1}{2}\|\nabla^2_{\bv\bv} f_n^\phi(\bw^*,\bv_n)\|. 
\end{array}
\end{eqnarray*}

\begin{proof} Since ${f_n^\varphi\geq0}$ and ${f_n^\phi\geq0}$, we have ${F\geq 0}$. This indicates $\Omega$ is bounded. As $f_n^\varphi$  is twice continuously differentiable, so is $g$ and thus gradient $\nabla  g$ and Hessian $\nabla^2 g$ are continuous, which together with the boundedness of $\Omega$ results in the boundedness of $\delta_{\bu}^*$ and thus $\tau^*$, that is ${\delta_{\bu}^*<\infty}$ and ${\tau^*<\infty}$. Similarly, we can prove that ${\delta_{\bv}^*<\infty}$ and ${\kappa^*<\infty}$.  

Now we prove the conclusion by induction. For $\ell=0$, direct calculation leads to the following chain of inequalities,
\begin{align*} 
\eqspace{1.5}
\begin{array}{lll}
&&\langle  {\boldsymbol\zeta}_n^{0}, \bu_n^1-{\bu}^* \rangle +    c_2 \|{\bu}_n^1-{\bu}^*\|^2 \\
&\leq&
\langle  {\boldsymbol\zeta}_n^{0}, \bu_n^1-{\bu}^* \rangle +  \tau^0 \|{\bu}^1_n-{\bu}_n^0\|^2+ c_2 \|{\bu}_n^1-{\bu}^*\|^2 \\
&\leq& \langle  {\boldsymbol\zeta}_n^{0}, \bu_n^0-{\bu}^* \rangle + c_2 \|{\bu}_n^0-{\bu}^*\|^2,
\end{array}
\end{align*}
where the second inequality is from \eqref{phase-iii-ADM-u}. Using the fact that $2\langle  \bu, \bv \rangle \leq  \|\bu\|^2/a+ a\|\bv\|^2$ for any $a>0$, the above condition  suffices to
\begin{align*} 
\eqspace{1.5}
\begin{array}{lll}
\frac{ c_2}{2} \|{\bu}_n^1-{\bu}^*\|^2 &\leq& \frac{1}{ c_2} \|{\boldsymbol\zeta}_n^{0}\|^2 +\frac{3 c_2}{2} \|{\bu}_n^0-{\bu}^*\|^2 \\
&\leq& \frac{1}{ c_2} \|{\boldsymbol\zeta}_n^{0}\|^2 +\frac{3}{2} F(\bu_n^0,\bv_n^0),
\end{array}
\end{align*}
resulting in 
\begin{align*} 
\eqspace{1.5}
\begin{array}{lll}\|{\bu}_n^1-{\bu}^*\|^2  \leq  \frac{2}{ c_2^2} \|{\boldsymbol\zeta}_n^{0}\|^2 +\frac{3}{ c_2} F(\bu_n^0,\bv_n^0) \leq\delta_{\bu}^*\end{array}
\end{align*}
 due to $(\bu_n^0,\bv_n^0)\in\Omega$, namely $ \bu_n^1\in \mathbb{B}(\bu^*,\delta_{\bu}^*)$. It is easy to see that  $ \bu_n^0\in \mathbb{B}(\bu^*,\delta_{\bu}^*)$. Therefore,   $\bu_n^0(a):=a\bu_n^0+(1-a)\bu_n^1\in \mathbb{B}(\bu^*,\delta_{\bu}^*)$ for any $a\in(0,1)$. Moreover, one can observe that $\bv_n^{0}\in\mathbb{B}(0,\frac{F^0}{ c_4})$ owing to $ c_4\|\bv_n^{0}\|^2\leq F^0$. Overall, by \eqref{def-tau-k}, we derive $\frac{1}{2}\|\nabla_{\bu\bu}^2 g(\bu_n^{0}(a),\bv_n^{0})\|\leq \tau^*$. Thus by the Mean Value Theorem and denoting $\triangle\bu^{1}_n:=\bu^{1}_n-\bu^{0}_n$ , we have 
\begin{align*} \label{strong-smooth-u}
\eqspace{1.5}
\begin{array}{lll}
&&g(\bu_n^{1},\bv_n^{0})- g(\bu_n^{0},\bv_n^{0}) -\langle {\boldsymbol\zeta}^{0}_n, \triangle\bu^{1}_n\rangle\\
&=& 
 \frac{1}{2}\langle \nabla_{\bu\bu}^2 g(\bu_n^{0}(a),\bv_n^{0})\cdot \triangle\bu^{1}_n , \triangle\bu^{1}_n\rangle \leq \tau^* \|\triangle\bu^{1}_n\|^2.
\end{array}
\end{align*}
Based on the above condition, we can obtain
\begin{align*} 
\eqspace{1.5}
\begin{array}{lll}
{F}(\bu_n^0,\bv_n^0)&=& 
g(\bu_n^0,\bv_n^{0}) + f^\phi_n(\bw^*,\bv_n^{0})\\
&+& c_2\|\bu_n^0-{\bu}^* \|^2+ c_4\|\bv_n^0\|^2\\
&\geq& g(\bu_n^{0},\bv_n^{0}) + f^\phi_n(\bw^*,\bv_n^{0})+ c_4\|\bv_n^0\|^2\\
&+&\langle {\boldsymbol\zeta}^{0}_n, \triangle\bu^{1}_n\rangle+ \tau^{0} \|\triangle\bu^{1}_n\|^2+  c_2\|\bu_n^1-{\bu}^* \|^2\\
&=&{F}(\bu_n^{1},\bv_n^{0}) + g(\bu_n^{0},\bv_n^{0}) - g(\bu_n^{1},\bv_n^{0})\\
&+&\langle {\boldsymbol\zeta}^{0}_n, \triangle\bu^{1}_n\rangle+ \tau^{0} \|\triangle\bu^{1}_n\|^2 \\
&\geq&  {F}(\bu_n^{1},\bv_n^{0})+(\tau^{0}-\tau^*) \|\triangle\bu^{1}_n\|^2 , 
\end{array}
\end{align*}
where the first inequality is from \eqref{phase-iii-ADM-u}. 

Similarly, it follows from \eqref{phase-iii-ADM-v} that 
\begin{eqnarray*} 
\eqspace{1.5}
\begin{array}{lll}
 && \langle {\boldsymbol\xi}^{0}_n, \bv_n^1\rangle  +  c_4 \|\bv_n^1\|^2 \\
&\leq & c_1 f^\varphi_n(\bu^{1},\bv_n^1) + \langle {\boldsymbol\xi}^{0}_n, \bv_n^1\rangle + \kappa^{\ell} \|\bv_n^1-\bv^{0}_n\|^2+  c_4 \|\bv_n^1\|^2 \\
&\leq & c_1 f^\varphi_n(\bu^{1},\bv_n^0) + \langle {\boldsymbol\xi}^{0}_n, \bv_n^0\rangle +  c_4 \|\bv_n^0\|^2 \\ 
&\leq &F(\bu^{1},\bv_n^0) + \langle {\boldsymbol\xi}^{0}_n, \bv_n^0\rangle, 
\end{array}
\end{eqnarray*} 
which results in
\begin{eqnarray*} 
\eqspace{1.5}
\begin{array}{lll}
    \frac{ c_4}{2} \|\bv_n^1\|^2 
&\leq &F(\bu^{1},\bv_n^0) +  \frac{1}{ c_4} \|{\boldsymbol\xi}^{0}_n\|^2+ \frac{ c_4}{2} \|\bv_n^0\|^2 \\
&\leq &\frac{3}{2}F(\bu^{1},\bv_n^0) +  \frac{1}{ c_4} \|{\boldsymbol\xi}^{0}_n\|^2\\
&\leq &\frac{3}{2}F(\bu^{0},\bv_n^0) +  \frac{1}{ c_4} \|{\boldsymbol\xi}^{0}_n\|^2. 
\end{array}
\end{eqnarray*} 
This condition gives rise to  
\begin{eqnarray*} 
\eqspace{1.5}
\begin{array}{lll}
  \|\bv_n^1\|^2  \leq  \frac{3}{ c_4}F(\bu^{0},\bv_n^0) +  \frac{2}{ c_4^2} \|{\boldsymbol\xi}^{0}_n\|^2\leq \delta_{\bv}^* 
\end{array}
\end{eqnarray*} 
 due to $(\bu_n^0,\bv_n^0)\in\Omega$, namely $ \bv_n^1\in \mathbb{B}(0,\delta_{\bv}^*)$. It is easy to see that  $ \bv_n^0\in \mathbb{B}(0,\delta_{\bv}^*)$. Therefore,   $\bv_n^0(a):=a\bv_n^0+(1-a)\bv_n^1\in \mathbb{B}(0,\delta_{\bv}^*)$ for any $a\in(0,1)$.  Overall, by the definition of $\kappa^*$, we derive $\frac{1}{2}\|\nabla_{\bv\bv}^2 f^\phi_n(\bw^*,\bv_n^0(a))\|\leq \kappa^*$. Thus by the Mean Value Theorem and denoting $\triangle\bv^{1}_n:=\bv^{1}_n-\bv^{0}_n$ , we have 
\begin{align*}
\eqspace{1.5}
\begin{array}{lll}
&& f^\phi_n(\bw^*,\bv_n^1)-  f^\phi_n(\bw^*,\bv_n^0) -\langle {\boldsymbol\xi}^{0}_n, \triangle\bv^{1}_n\rangle\\
&=& 
 \frac{1}{2}\langle \nabla_{\bv\bv}^2 f^\phi_n(\bw^*,\bv_n^0(a))\cdot \triangle\bv^{1}_n , \triangle\bv^{1}_n\rangle \leq \kappa^* \|\triangle\bv^{1}_n\|^2.
\end{array}
\end{align*}
The above condition enables us to derive 
\begin{align*} 
\eqspace{1.5}
\begin{array}{lll}
&&{F}(\bu_n^1,\bv_n^0)\\
&=&  g(\bu_n^1,\bv_n^0) +   f^\phi_n(\bw^*,\bv_n^0)+ c_2 \|{\bu^1_n}-{\bu}^* \|^2 + c_4\|\bv_n^0\|^2 \\
&\geq& g(\bu_n^1,\bv_n^0) +    c_2 \|{\bu^1_n}-{\bu}^* \|^2 + c_4\|\bv_n^0\|^2 \\
&+& f^\phi_n(\bw^*,\bv_n^1) -\langle {\boldsymbol\xi}^{0}_n, \triangle\bv^{1}_n\rangle- \kappa^* \|\triangle\bv^{1}_n\|^2 \\
&\geq& g(\bu_n^1,\bv_n^1)+f^\phi_n(\bw^*,\bv_n^1) +  c_2\left\|{\bu^1_n}-{\bu}^*\right\|^2 \\
&+&  c_4\|\bv_n^1\|^2+(\kappa^0- \kappa^*) \|\triangle\bv^{1}_n\|^2 \\
&=&{F}(\bu_n^{1},\bv_n^{1})+(\kappa^0- \kappa^*) \|\triangle\bv^{1}_n\|^2 ,
\end{array}
\end{align*}
where the second inequality is from  \eqref{phase-iii-ADM-v}.  Therefore, we prove 
\begin{align*} 
\eqspace{1.5}
\begin{array}{lll}
&&{F}(\bu_n^0,\bv_n^0)\\
&\geq&  {F}(\bu_n^{1},\bv_n^{0})+(\tau^{0}-\tau^*) \|\triangle \bu^{1}_n\|^2\\
&\geq & {F}(\bu_n^{1},\bv_n^{1})+(\tau^{0}-\tau^*) \|\triangle\bu^{1}_n \|^2 +(\kappa^0- \kappa^*) \|\triangle\bv^{1}_n\|^2.
\end{array}
\end{align*}
This indicates ${F}(\bu_n^{1},\bv_n^{1})\leq {F}(\bu_n^0,\bv_n^0)=F^0$ and hence $(\bu_n^{1},\bv_n^{1})\in\Omega$. Repeating the above proof, we can show that 
\begin{align*} 
\eqspace{1.5}
\begin{array}{lll}
&&{F}(\bu_n^1,\bv_n^1)
- {F}(\bu_n^{2},\bv_n^{2})\\
&\geq& (\tau^{1}-\tau^*) \|\triangle \bu^{2}_n \|^2 +(\kappa^1- \kappa^*) \|\triangle\bv^{2}_n\|^2,
\end{array}
\end{align*}
resulting in $(\bu_n^{2},\bv_n^{2})\in\Omega$. Then repeating the proof enables us to prove \begin{align*} 
\eqspace{1.5}
\begin{array}{lll}
&&{F}(\bu_n^{\ell},\bv_n^{\ell})
- {F}(\bu_n^{\ell+1},\bv_n^{\ell+1})\\
&\geq& (\tau^{\ell}-\tau^*) \|\triangle \bu^{\ell+1}_n \|^2 +(\kappa^{\ell}- \kappa^*) \|\triangle\bv^{\ell+1}_n\|^2,
\end{array}
\end{align*}
for any ${\ell\geq0}$. Hence sequence $\{{F}(\bu_n^{\ell},\bv_n^{\ell})\}$ is strictly decreasing, which together with ${F (\bu_n^{\ell},\bv_n^{\ell})\geq0}$ ensures its convergence. Finally taking the limit of both sides of the above condition leads to ${\triangle \bu^{\ell+1}_n \to 0}$ and ${\triangle\bv^{\ell+1}_n\to 0}$ when $\ell\to \infty.$ The proof is finished.
\end{proof}

\end{document}